\documentclass[a4paper,10pt]{article}

\usepackage{cite, amsfonts, amsthm,  fullpage}
\usepackage{amssymb}
\usepackage{amsbsy}
\usepackage{epsfig}
\usepackage{verbatim}

\newcommand{\Pf}{\mathop\mathrm{Pf}\nolimits}
\newcommand{\sgn}{\mathop\mathrm{sgn}\nolimits}

\newcommand{\bt}{\mathbf{t}}

\theoremstyle{plain}

\newtheorem{Lemma}{Lemma}
\newtheorem{Proposition}{Proposition}
\newtheorem{corollary}{Corollary}

\theoremstyle{remark}
\newtheorem{Remark}{Remark}

\def\l{\langle}
\def\r{\rangle}

\def\det{\mathrm {det}}

\def\res{\mathop{\mathrm {res}}\limits_}

\def\bp{\begin{Proposition}\rm}
\def\ep{\end{Proposition}}
\def\bc{\begin{corollary}}
\def\ec{\end{corollary}}
\def\bl{\begin{Lemma}\em}
\def\el{\end{Lemma}}
\def\be{\begin{equation}}
\def\ee{\end{equation}}
\def\br{\begin{Remark}\rm\small}
\def\er{\end{Remark}}
\def\brs{\begin{remarks}.\\ \rm\
\begin{enumerate}}
\def\ers{\end{enumerate}\end{remarks}}
\def\bea{\begin{eqnarray}}
\def\eea{\end{eqnarray}}

\def\det{\mathrm {det}}

\def\sgn{\mathrm {sgn}}

\def\res{\mathop{\mathrm {res}}\limits}

\def\&{&{\hskip -20pt}}

\newcount\YDcount\YDcount=0
\def\YDsize{10pt}

\def\YD#1{%
\ifnum#1=0
 \ifnum\YDcount=0 \ifx\varnothing\undefined\emptyset\else\varnothing\fi
 \else\vskip1.4pt\egroup\YDcount=0\fi
\else
 \ifnum\YDcount=0 \YDcount=1\vcenter\bgroup\vskip1pt
 \else\nointerlineskip\fi
 \vbox{\hrule\hbox{\vrule height\YDsize
 \loop\hskip\YDsize\vrule\ifnum\YDcount<#1\advance\YDcount1\repeat}\hrule
 \kern-0.4pt}\expandafter\YD
\fi}

\begin{document}
\author{ J. W. van de Leur\thanks{Mathematical Institute, Utrecht University, P.O. Box 80010, 3508 TA Utrecht, 
The Netherlands, email: J.W.vandeLeur@uu.nl}\ \and A. Yu.
Orlov\thanks{Institute of Oceanology RAS, Nahimovskii Prospekt 36,
Moscow, Russia, and National Research University Higher School of Economics, 
International Laboratory of Representation 
Theory and Mathematical Physics,
20 Myasnitskaya Ulitsa, Moscow 101000, Russia, email: orlovs@ocean.ru}}
\title{Pfaffian and determinantal tau functions I}

\maketitle

\begin{abstract}

Adler, Shiota and van Moerbeke observed that a tau function of 
the Pfaff lattice is a square root of a tau function of the  Toda lattice hierarchy of  Ueno and  Takasaki. In this paper we give a representation theoretical explanation for this phenomenon.
We consider 2-BKP and two-component 2-KP tau functions. We shall show
that a square of a BKP tau function is equal to a certain two-component KP tau function 
and a square of a 2-BKP tau function is equal to a certain two-component 2-KP tau function.

\end{abstract}

\bigskip 


\section{Introduction}

Sato and Sato\cite{Sato} and Date, Jimbo, Kashiwara and Miwa \cite{DJKM} introduced and described in the beginning of the 1980's 
the KP hierarchy in various setting. They introduced a tau function, which is a fundamental object in this theory. It is an 
element in a $GL_\infty$ group orbit and as such a solution of the KP hierarchy. Around the same time  Date, Jimbo, 
Kashiwara and Miwa introduced in \cite{DJKM-BKP} also a new hierarchy of soliton equations, which they called the KP hierarchy 
of type B or BKP hierarchy. The corresponding tau function is an element of the $B_\infty$ group orbit and hence a solution to 
this new BKP hierarchy. In a straightforward calculation they show that this BKP tau function is the square root of a certain KP 
tau function. Their proof of this phenomenon is fundamental for the contents of this paper. Here we show that this method is 
also  applicable for the observation of Adler, Shiota and van Moerbeke \cite{AvM-Pfaff}, \cite{AMS}, that a tau function of 
the Pfaff lattice is a square root of an Ueno, Takasaki \cite{UT} Toda lattice tau function.
In this introduction we will recall the work of Date, Jimbo, Kashiwara and Miwa \cite{DJKM} and explain the relation BKP versus KP.
The involutions that are used in their work, which provide the relation between the two tau functions of KP and BKP, also give 
the relation between the various tau functions of the Pfaff and Toda lattice. This means that the construction of the KP group element out of the BKP group element is the same. One finds in both cases that $g_{\rm KP}=h_{\rm BKP}\hat h_{\rm BKP}$, where $\hat h_{\rm BKP}$ is constructed out of $h_{\rm BKP}$ by using one of the involutions.

The Pfaff lattice of Adler and van Moerbeke \cite{AvM-Pfaff}, \cite{AMS} was discovered by Jimbo and Miwa (\cite{JM}, section 7) 
it was rediscovered by Hirota and Ohta \cite {HO} as  the coupled KP hierarchy and studied in a paper by Kac and one of the 
authors \cite{KvdLbispec} as the charged DKP hierarchy. The Pfaff lattice studied in \cite{L1} is slightly bigger than the one 
studied by Adler, Shiota and van Moerbeke in  \cite{AMS} it is the charged BKP hierarchy of \cite{KvdLbispec}, called the large BKP 
in \cite{OST-I} to make difference to the small BKP of \cite{DJKM-BKP}.
Here we study this charged or large BKP.

In addition we also consider the large 2-BKP which is two-sided evolution (via positive and negative parts of current modes,
like the Toda lattice hierarchy of Ueno and Takasaki \cite{UT} in contrast to the one-sided KP). There we have two sets of higher
times, $t$ and ${\bar t}$. This is the B-analogue 
of Takasaki's 2-DKP of \cite{T-09} (which he called the Pfaff lattice by
analogy with Pfaff lattice of Adler and van Moerbeke).
Also we use two discrete variables for BKP (and also for 2-BKP): $\tau_{m,n}(t,{\bar t})$. 

In this introduction and the first 5 sections  we will only describe the one sided case, to avoid technicalities. In section 6 
we will introduce the 2-sided case, which will be important for matrix models.

For matrix models we need the semi-infinite Toda and Pfaff lattices. The
fermionic expressions for the related tau functions is considered in the Section 7 where a set of examples
is written down. In this Section we generalize the observation of Adler, van Moerbeke and Shiota \cite{AvM-Pfaff},
\cite{AMS} about the relation between Toda and Pfaff lattices.

But first we will start with the recollection of the work of Date, Jimbo, Kashiwara, and Miwa, on KP and BKP hierarchies.

\paragraph{Fermions and KP hierarchy}
Consider, following \cite{DJKM} and  \cite{JM} charged fermions $\{ f_i,f^\dag _i\}_{i\in\mathbb{Z}}$, satisfying
 \be
 [f_i,f_j^{\dag} ]_+=\delta_{ij}\,,\quad [f_i,f_j]_+ = 0 =
 [f_i^{\dag},f_j^{\dag}]_+ \,,\quad i,j \in\mathbb{Z}
 \ee
 where $[x,y]_+ =xy+yx$ is anticommutator. The elements $f_i$ and $f_i^\dag$ form the basis of a vector space, which we denote 
 by $V$. 
Introduce the spin modules (right and left Fock spaces) with vacuum vectors, $|0\rangle $, $ \langle 0|$, such that
  \be
  f_i|0\rangle =f_{-1-i}^{\dag}|0\rangle =0= \langle 0|f_i^{\dag} =\langle 0|f_{-1-i}\,,\quad i<0
  \ee
  Let $g \in GL_\infty$ which may be written as
   \be
   g=\exp \sum_{i,j\in\mathbb{Z}} a_{ij}:f_if_j^{\dag}:
   \ee
 where $:f_if_j^{\dag}:$   denotes $f_if_j^{\dag}-\langle 0|f_if_j^{\dag}|0\rangle$, and $a_{ij}$ are some complex numbers. 
 The elements $:f_if^{\dag}_j:$ together with 1 form a basis of the Lie algebra $gl_\infty$, see \cite{DJKM}, \cite{JM} 
 for more details.
 
  Then
  \be\label{KPHirota}
  \sum_{i\in\mathbb{Z}}\, f_ig|0\rangle \otimes f_i^{\dag}g|0\rangle =0
  \ee
  is the KP hierarchy in the fermionic form, see \cite{DJKM}.

One turns this equation into a hierarchy of differential equation by using the boson-fermion correspondence, see e.g. 
\cite{KvdL} for more details. 

However the tau function can be be calculated in a different way, define the oscillator algebra 
  \be\label{foscillator}V=
  \{  \alpha_n = \sum_{i \in\mathbb{Z}} :f_i f_{i+n}^{\dag}:\} 
  \ee
  We have
  \be
  [\alpha_n,\alpha_m]_- = n\delta_{n+m,0}
  \ee
 where $[x,y]_- =xy-yx$ is commutator.
Define
\be
\label{fz}
f(z)=\sum_{i\in\mathbb{Z}} f_iz^i, \qquad
f^\dagger(z)=\sum_{i\in\mathbb{Z}} f_i^\dagger z^{-i}\, ,
\ee
then 
\be
\label{fz-alpha}
[\alpha_k, f(z)]=z^k f(z)\quad\mbox{and }  [\alpha_k, f^\dagger (z)]=-z^k f^\dagger (z)]
\ee
Now
\be 
\alpha_n|0\rangle =\langle 0|\alpha_{-n}=0,\qquad n\ge 0\, . 
\ee
Then $\tau_{\rm KP}(t)$, defined by  the following expectation value
 \be
\tau_{\rm KP}(t)=\langle 0 |\exp\left( \sum_{k=1}^\infty t_k\alpha_k \right)g|0\rangle\, ,
\ee
is a solution of the KP hierarchy.

\paragraph{Neutral fermions and the BKP hierarchy. \label{neutralBKP}}

Following \cite{JM} we introduce  an involution $\omega$ on the Clifford algebra, defined by
  \be
\label{omega}
  \omega(f_i)=(-)^if_{-i}^{\dag}\,,\quad \omega(f_i^{\dag})=(-)^if_{-i}
  \ee
The fixed points of $\omega$ in the vector space $V$ with basis $f_i, f^{\dag}_i$  are  the elements
\be\label{neutral-fermions}
b_j=\frac{f_j+(-)^jf_{-j}^{\dag}}{\sqrt 2}\, .
\ee
The elements
\be\label{neutral-fermions-hat}
\hat b_j=i\frac{f_j-(-)^jf_{-j}^{\dag}}{\sqrt 2}
\ee 
are elements in the $-1$ eigenspace of $\omega$.
$b_j$ and $\hat b_j$ form a new basis of $V$
We have
\be\label{neutral-fermions-anticomm}
[b_i,\hat b_j]=0, \quad [b_i, b_j]=[\hat b_i,\hat b_j]=(-)^i\delta_{i,-j}\, .
\ee
The fixed points in $gl_\infty$ are 1 and the elements
\be
(-)^k:f_jf^{\dag}_{-k}:-(-)^{j}:f_{k}f_{-j}^{\dag}:
\, .
\ee
These elements form a Lie algebra of type $b_\infty$. If one considers the action of these elements on the vacua of the 
spin modules one obtains a level two representation of $b_\infty$ (Note that here we allow also certain infinite sums of 
these elements).

Note also that only the $\alpha_k$ for $k$ odd are fixed by $\omega$.

It is straightforward to check that
\be
(-)^k:f_jf^{\dag}_{-k}:-(-)^{j}:f_{k}f_{-j}^{\dag}:
=:b_jb_{k}:+:\hat b_j\hat b_{k}:\,.
\ee
The following observation is crucial. The elements $:b_jb_{k}:$ together with 1, or the elements $:\hat b_j\hat b_{k}:$ 
together with 1, separately  also form the Lie algebra $b_\infty$. If one considers the action of these elements separately 
on the vacua one obtains a level one representation. The corresponding module in terms of the $b_j$ or $\hat b_j$ is called 
the spin module of type B.
One has
\be\label{b-vacua}
b_j|0\rangle=\hat b_j|0\rangle=\langle 0|b_{-j}=\langle 0|\hat b_{-j}, \quad j<0.
\ee
and
\be\label{b-0-vacua}
b_0|0\rangle=-i\hat b_0|0\rangle=\frac12 \sqrt 2 f_0|0\rangle,
\quad\mbox{and }
\langle 0|b_0=i\langle 0|\hat b_0
=\frac12 \sqrt 2 \langle 0|f^{\dag}_0\,
,
\ee
hence
\be\label{b-0-b-0-exp-value}
\langle 0|b_0b_0|0\rangle=-i\langle 0|b_0\hat b_0|0\rangle=i\langle 0|\hat b_0 b_0|0\rangle=
\langle 0|\hat b_0\hat b_0|0\rangle=\frac12\, .
\ee
As in the KP case we now consider an element in the group $h\in B_\infty$, 
such an element may e.g. be written as
\be
\label{h}
   h=\exp \sum_{i,j\in\mathbb{Z}} a_{ij}:b_ib_j:
   \ee
Then 
 \be\label{BKPHirota}
  \sum_{i\in\mathbb{Z}}(-)^i\, b_i h|0\rangle \otimes b_{-i}h|0\rangle =hb_0|0\rangle \otimes h b_{0}|0\rangle
  \ee
  is the BKP hierarchy in the fermionic form, see e.g. \cite{JM}.
One turns this into a hierarchy of differential equations by using twisted vertex operators see \cite{KvdLbispec}
for more details. The corresponding BKP tau function can be obtained as follows.
Define the twisted oscillator algebra 
\be\label{boscillator}
  \gamma_n = \frac12 \sum_{i \in\mathbb{Z}} (-)^{i+1}:b_i b_{-i-n}:\, .
  \ee
Then $\gamma_n=0$ if $n$ is even. We have
  \be\label{Heisenberg-gamma}
  [\gamma_n,\gamma_m]_- = \frac{n}{2}\delta_{n+m,0}
  \ee
 and
\be\label{gamma-vac} 
\gamma_n|0\rangle =\langle 0|\gamma_{-n}=0,\qquad n\ge 0\, . 
\ee
Let 
\be
\label{bz}
b(z)=\sum_{i\in\mathbb{Z} } b_i z^i,
\ee
then 
\be 
\label{bz-gamma}
[\gamma_k,b(z)]=z^k b(z).
\ee
The BKP tau function $\tau_{\rm BKP}(t)$ is defined by  the following expectation value
 \be
\tau_{\rm BKP}(t)=\langle 0 |\exp\left( \sum_{k=1,\  {\rm odd}}^\infty t_k\gamma_k \right)h|0\rangle
\ee
and is a solution of the BKP hierarchy.
A crucial observation is the fact that in this construction we could have replaced the $b_n$ by $\hat b_n$ and in this 
way would have obtained the same result.

\paragraph{A relation between KP and BKP tau functions}
 We follow Jimbo and Miwa \cite{JM} or rather You \cite{Y} and define an automorphism $\hat{}$ on the B type
 Clifford algebra
  \be
\label{hat}
  \hat{}\,   (b_n)={\hat b_n}
  \ee
then 
\be
\label{bb}
:b_nb_m:+\hat{}\,   (:b_nb_m:)=:b_nb_m:+:\hat b_n\hat b_m:=(-)^m:f_nf_{-m}^{\dag}:-(-)^n:f_mf_{-n}^{\dag}:
\ee
and
\be
\label{obs1}
\gamma_n+ \hat{}\,   (\gamma_n)=\gamma_n+\hat\gamma_n=\alpha_n,\quad n\ \mbox odd.
\ee
Then
\be
\left( \tau_{\rm BKP}(t)\right)^2=
\langle 0 |\exp\left( \sum_{k=1,\  {\rm odd}}^\infty t_k\gamma_k \right)h|0\rangle
\langle 0 |\exp\left( \sum_{k=1,\  {\rm odd}}^\infty t_k\hat\gamma_k \right)\hat h|0\rangle 
\ee
and since the elements $h$ and  $\gamma_n$ commute with $\hat h$ and  $\hat \gamma_n$
\be
\left( \tau_{\rm BKP}(t)\right)^2=
\langle 0 |\exp\left( \sum_{k=1,\  {\rm odd}}^\infty t_k\left(\gamma_k +\hat\gamma_k \right)\right) h\hat h|0\rangle \, ,
\ee
Now using (\ref{obs1}), we find  
\be
\left( \tau_{\rm BKP}(t)\right)^2
=\langle 0 |\exp\left( \sum_{k=1,\  { odd}}^\infty t_k\alpha_k \right) h\hat h|0\rangle =
\tau_{\rm KP}(t_1,0,t_3,0,t_5,0,\ldots),
\ee
for $g=h\hat h$.
Since $h$ is of the form (\ref{h}) we find, using (\ref{bb}), that
\be
g=h\hat h= \exp \sum_{n,m\in\mathbb{Z}} a_{nm}(:b_nb_m:+:\hat b_n \hat b_m:)
=\exp \sum_{n,m\in\mathbb{Z}} a_{nm}((-)^m:f_nf_{-m}^{\dag}:-(-)^n:f_mf_{-n}^{\dag}:)\, .
\ee 
The element $(-)^m:f_nf_{-m}^{\dag}:-(-)^n:f_mf_{-n}^{\dag}:$ is fixed by $\omega$,
hence $g\in GL_\infty$ is an element in the level 2 representation of $b_\infty$, which one gets by considering the action 
of the $\omega$ invariant elements in $gl_\infty$. In other words this $\tau_{\rm KP}(t_1,0,t_3,0,t_5,0,\ldots)$ is an 
element in the $B_\infty$ group orbit of the vacuum, where we consider $B_\infty$ as a subgroup of $GL_\infty$ and not as a 
group acting on the spin module of type B. The latter one is related to the level 1 representation of $b_\infty$.

\paragraph{A relation between KP and BKP wave functions}
Recall from (\ref{fz}) the formula for $f(z)$, the KP wave function is defined as 
$W_{\rm KP}(t, z)=\frac{w_{\rm KP}(t,z)}{\tau_{\rm KP}(t)}$,
with
\be
w_{\rm KP}(t,z)=\langle 0|f_0^\dagger\exp\left( \sum_{k=1}^\infty t_k\alpha_k \right) f(z)g|0\rangle\, .
\ee
Using the boson-fermion correspondence (see e.g. \cite{DJKM}, \cite{KvdL} or (\ref{THETA}) in the next section),
\[
  f(z)=Q z^{\alpha_0}e^{-\sum_{k<0}-\frac{\alpha_k}{k}} z^{-k} e^{-\sum_{k>0}\frac{\alpha_k}{k} z^{-k}}\, , 
\]
which is based on (\ref{fz-alpha}), then gives
 \be
w_{\rm KP}(t,z)=
e^{\sum_{k>0} t_kz^k} \langle 0|\exp\left( \sum_{k=1}^\infty \left(t_k-\frac{z^{-k}}{k}\right)\alpha_k \right) g|0\rangle
\ee
\be 
=\tau_{\rm KP}(t_1-\frac{z^{-1}}{1},t_2-\frac{z^{-2}}{2},t_3-\frac{z^{-3}}{3},\ldots)e^{\sum_{k>0} t_kz^k}
\ee

Recall $b(z)$ from (\ref{bz}) and define analogously  $\hat b(z)=\sum_{j\in\mathbb{Z}} \hat b_jz^j$, then 
using (\ref{neutral-fermions}) and (\ref{neutral-fermions-hat})
\be
f(z)= \frac{1}{\sqrt 2}\left(b(z)-i\hat b(z)\right)\quad\mbox{and }
f^\dagger (z)= \frac{1}{\sqrt 2}\left(b(-z)+i\hat b(-z)\right)
\ee
Now let as before $g=h\hat h$, then 
\be
w_{\rm KP}(t_{\rm odd},z)=
w_{\rm KP}(t_1,0,t_3,0,\ldots,z)=
\tau_{\rm KP}(t_1-\frac{z^{-1}}{1},-\frac{z^{-2}}{2},t_3-\frac{z^{-3}}{3},-\frac{z^{-4}}{4},\ldots)
e^{\sum_{{k>0}\  { odd}} t_kz^k}
\ee
Using (\ref{bz-gamma}) and the known bosonization formula (6.5) of \cite{JM}, 
\[
 \l 0|b_0 b(z)e^{\sum_{k=1,\,{\rm odd}} \gamma_k t_k}=\frac12
 \l 0| e^{\sum_{k=1,\,{\rm odd}} \gamma_k \left(t_k -\frac{2z^{-k}}{k} \right)}
\]
and a similar formula with ``hats'' ((7.4) in  \cite{JM}),
 we obtain
\be
\label{wave-KP} 
w_{\rm KP}(t_{\rm odd},z)=\frac12 \langle 0|\left(b_0+i\hat b_0\right)
\exp\left( \sum_{k=1,\  {\rm odd}}^\infty t_k\left(\gamma_k +\hat\gamma_k\right)\right)
\left(b(z)-i\hat b(z)\right)h\hat h|0\rangle
\ee
\[
 =\frac12e^{\sum_{k>0\ odd} t_kz^k} 
\langle 0|\exp\left( \sum_{k=1,\  {\rm odd}}^\infty \left(t_k-2\frac{z^{-k}}{k}\right)\gamma_k\right)h|0\rangle
\langle 0|\exp\left( \sum_{k=1,\  {\rm odd}}^\infty t_k\hat\gamma_k\right)
\hat h|0\rangle
\]
 \[
  +\frac12e^{\sum_{k>0\ odd} t_kz^k} 
\langle 0|\exp\left( \sum_{k=1,\  {\rm odd}}^\infty t_k\gamma_k\right)h|0\rangle
\langle 0|\exp\left( \sum_{k=1,\  {\rm odd}}^\infty \left(t_k-2\frac{z^{-k}}{k}\right)\hat\gamma_k\right)
\hat h|0\rangle
 \]
\[
 =\tau_{\rm BKP}(t)\tau_{\rm BKP}(t_1-2\frac{z^{-1}}{1},t_3-2\frac{z^{-3}}{3},\ldots)e^{\sum_{k>0\ odd} t_kz^k} \, .
\]
If we now divide this by 
\[
\tau_{\rm KP}(t_1,0,t_3,0,\ldots)=(\tau_{\rm BKP}(t))^2\, , 
\]
we obtain  an expression for $W_{\rm KP}(t_{\rm odd},z)$.
Define
\be\label{bkp-wave-function}
W_{\rm BKP}(t,z)=2\frac{\l 0|b_0 e^{\sum_{k=1,\,{\rm odd}} t_k\gamma_k}b(z)  h  |0 \r}
{\l 0| e^{\sum_{k=1,\,{\rm odd}} t_k\gamma_k} h  |0\r} = e^{\sum_{k>0,\,{\rm odd}} t_kz^k}\left(1+O(z^{-1})\right)\, .
\ee
Again using (\ref{bz-gamma}), one deduces
\be
\label{bkp-wave-function2}
W_{\rm BKP}(t,z)=\tau_{\rm BKP}(t_1-2\frac{z^{-1}}{1},t_3-2\frac{z^{-3}}{3},\ldots)e^{\sum_{k>0\ odd} t_kz^k}/
\tau_{\rm BKP}(t)\, 
\ee
Now comparing (\ref{wave-KP}) and (\ref{bkp-wave-function2}), we obtain:
\be
W_{\rm BKP}(t,z)=W_{\rm KP}(t_1,0,t_3,0,\ldots,z)\, .
\ee

\section{Another realization of the KP hierarchy}
In order to describe the relation between the Toda lattice hierarchy (or 2-component KP) and the Pfaff lattice 
(or large BKP), we need another realization of both spin modules. We start with the KP hierarchy and 
relabel the fermions $f_i$ and $f_i^\dag$ as follows
 \be\label{relabel-1}
 f_0 = \psi \,,\quad f_0^\dag=\psi^\dag
 \ee
  \be
 f_{2i+1}= \psi_i^{(1)}\,,\quad f_{2i+1}^\dag= \psi_i^{\dag(1)}\,,\quad i \ge 0
 \ee
 while
  \be
 f_{2i+2}= \psi_i^{(2)}\,,\quad f_{2i+2}^\dag= \psi_i^{\dag(2)}\,,\quad i\ge 0
 \ee
  \be
 f_{2i}= \psi_i^{(1)}\,,\quad f_{2i}^\dag= \psi_i^{\dag(1)}\,,\quad i<0
 \ee
  \be
\label{above}
 f_{2i+1}= -\psi_i^{(2)}\,,\quad f_{2i+1}^\dag= -\psi_i^{\dag(2)}\,,\quad i<0.
 \ee
 The minus sign in (\ref{above}) will be convenient later on when we apply the involution $\omega$ to this new realization.
 Then for $a,b=1,2$
   \be
  [\psi_i^{(a)},\psi_j^{\dag(b)}]_+ =\delta_{a,b}\delta_{i,j},\qquad \mbox{and}\qquad 
  [\psi ,  \psi^\dag]_+ =1\,,
  \ee
all other anti-commutation relations are zero.
The action on the vacuum is given by
  \be\label{psi-vac}
  \psi^\dag|0\rangle =0 = \langle 0|\psi
  \ee
  \be
  \psi_i^{(a)}|0\rangle = \psi_{-1-i}^{\dag(\alpha)}|0\rangle =0
  =\langle 0|\psi_i^{\dag(\alpha)}=\langle 0|\psi_{-1-i}^{(a)}\,,\quad i<0
  \ee 
 Now the KP equation (\ref{KPHirota}) may be rewritten as
  \be\label{KPHirota2}
  \psi g|0\rangle \otimes \psi^\dag g|0\rangle
  +\sum_{a=1,2}\sum_{i\in\mathbb{Z}}\, \psi^{(a)}_ig|0\rangle \otimes \psi_i^{\dag(\alpha)}g|0\rangle =0
  \ee
 where $g$ is the same as in (\ref{KPHirota}) rewritten as
  \be
  g=e^{\sum_{a,b=1,2}\sum_{n,m\in\mathbb{Z}}a_{nm}^{ab}\psi^{(a)}_n \psi^{\dag(b)}_m +
   \sum_{a=1,2}\sum_{n\in\mathbb{Z}} a_{nm}^{a0}\psi\psi^{\dag(\alpha)}_n +
  \sum_{a=1,2}\sum_{n\in\mathbb{Z}}a_{nm}^{0a}\psi^{(a)}_n \psi^\dag + a^{00}\psi\psi^\dag} . 
  \ee
  \br
  One can identify eq. (\ref{KPHirota2}) with the Hirota equation for the three-component KP hierarchy 
  \cite{DJKM},\cite{JM},\cite{KvdL}
  restricted on the class of $g$ which does not depend on $\psi^{(3)}_k,\psi^{\dag(3)}_k$ for all $k \neq i$
  except on $\psi^{(3)}_i:=\psi,\psi^{\dag(3)}_i:=\psi^\dag$ where $i$ is arbitrary chosen.
    
  \er

  It may also be written in the form
  \be\label{developement}
  g=g^{(0)}+g^{(1)}\psi + g^{(-1)}\psi^\dag\,,\quad
  g^{(0)}=g^{(0,0)}+g^{(1,-1)}\psi\psi^\dag
  \ee
  where $g^{(0,0)}$, $g^{(1,-1)}$, $g^{(1)}$ and 
  $g^{(-1)}$ do not contain neither $\psi$, nor $\psi^\dag$. Then,
  $g^{(0)}$ is even and $g^{(\pm 1)}$ are odd in the $Z_2$ grading of the Clifford algebra.

 Define a 2-component oscillator algebra (cf (\ref{foscillator}))
  \be\label{oscillator}
  \alpha^{(a)}_n = \sum_{i \in\mathbb{Z}} :\psi_i^{(a)} \psi_{i+n}^{\dag(a)}:\, .
  \ee
  We have
  \be\label{Heisenberg-alpha}
  [\alpha^{(a)}_n,\alpha^{(b)}_m]_- = n\delta_{a,b}\delta_{n+m,0}
  \ee

 Introduce
 \be \label{Psi-a}
\Psi^{(a)}_n =
\cases{
\psi^{(a)}_{n-1}\cdots \psi^{(a)}_{0} \, \quad {\rm if }\quad n > 0 \cr
1 \qquad\qquad \, \qquad \,  {\rm if }\quad n = 0 \cr
\psi^{\dag(a)}_{n}\cdots \psi^{\dag(a)}_{-1} \quad \,  {\rm if }\quad n < 0 }
    \\ 
 \,,\quad
\Psi^{\dag(a)}_n =
\cases{
\psi^{\dag(a)}_{0}\cdots \psi^{\dag(a)}_{n-1} \, \quad {\rm if }\quad n > 0 \cr
1 \qquad\qquad \, \qquad \,  \,{\rm if }\quad n = 0 \cr
\psi^{(a)}_{-1}\cdots \psi^{(a)}_{n} \quad \quad {\rm if }\quad n < 0 }
 \ee
 and introduce right and left Fock vectors 
  \be\label{2-vacuum-0}
  |n,m,0\r = \Psi^{(2)}_m \Psi^{(1)}_n |0\r\,,\quad \l n,m,0| = \l 0|\Psi^{\dag(1)}_n \Psi^{\dag(2)}_m 
  \ee
  \be\label{2-vacuum-1}
  |n,m,1\r = \Psi^{(2)}_m \Psi^{(1)}_n \psi |0\r\,,\quad 
  \l n,m,1| = \l 0|\psi^\dag \Psi^{\dag(1)}_n \Psi^{\dag(2)}_m 
  \ee
  which are highest weight vectors for the oscillator algebra (\ref{oscillator}):
  \be\label{highest-weight-OA}
  \alpha_i^{(a)}|n,m,k\r = 0 \quad {\rm if}\quad i>0\,, \qquad \l n,m,k| \alpha_{-i}^{(a)}=0 \quad {\rm if}\quad i>0
  \ee
  for any $a,n,m$ and $k=0,1$.
  
  We have the orthogonality condition $\l n,m,k|1|n',m',k'\r = \delta_{n,n'}\delta_{m,m'}\delta_{k,k'}$ where
  $n,n',m,m' \in\mathbb{Z}$, $k,k'=0,1$.
  
  Now we can write
  \be\label{g-vac}
  g|0\r = \sum_{n\in\mathbb{Z}}\left(g^{(0)}_{-n}|n,-n,0\r + g^{(1)}_{-n}|n-1,-n,1\r \right) 
  \ee

  \paragraph{Boson-fermion correspondence.}
  
  Now we set the following correspondence. The right Fock space is isomorphic to
$\mathbb[\theta, q_a, q_a^{-1}, t_i^{(a)}; \, a=1,2,\,  i=1,2,\ldots]$. Let $\sigma$ be the corresponding isomorphism. 
Introduce the fermionic fields
 \be\label{FF}
\psi^{(a)}(z)=\sum_{i\in \mathbb{Z}}\psi_i^{(a)} z^i,\qquad 
\psi^{\dagger(a)}(z)=\sum_{i\in \mathbb{Z}}\psi_i^{\dagger(a)} z^{-i-1}\, , 
 \ee
then 
  \be
\label{THETA}
  \sigma(\psi)=\theta\,,\quad \sigma(\psi^\dag)=\frac{\partial}{\partial \theta}
  \ee
  where $\theta^2=0$, and
  \be
  \sigma(\psi^{(a)}(z))=q_a z^{q_a\frac{\partial}{\partial q_a}} E^{+(a)}(z)\,,\quad
  \sigma(\psi^{\dag(a)}(z))=q_a^{-1} z^{-q_a\frac{\partial}{\partial q_a}} E^{-(a)}(z)
  \ee
  where
  \be
  E^{\pm(a)}(z)= e^{ \pm\sum_{n=1}^\infty t_n^{(a)} z^n } 
  e^{ \mp\sum_{k=1}^\infty \frac 1k \frac{\partial}{\partial t_n^{(a)}}} 
  \ee
Here
  \be\label{anticomm-q}
 q_aq_b=-q_bq_a\,,\quad \theta q_a=-q_a\theta 
  \ee
for $a,b=1,2$ and $a\neq b$.
  
  Then (\ref{g-vac}) turns out to be
  \be
 \sigma(g|0\r )= \sum_{n} \left( \tau_{-n}^{(0)} q_2^{n}q_1^{-n} + \tau_{-n}^{(1)} q_2^{n}q_1^{-n-1} \theta\right)
  \ee
  
  Now the Hirota KP equations (\ref{KPHirota}) 
  (rewritten in form (\ref{KPHirota2}))  are
  \[
  0=-\sum_{n,m} \tau_{-n}^{(0)} q_2^{n}q_1^{-n}\theta \otimes   \tau_{-m}^{(1)} q_2^{m}q_1^{-m-1} +
  \]
  \[
   \res_z (-)^n z^{-n}E^{+(1)}(z)\left(\tau_{-n}^{(0)} q_2^{n}q_1^{1-n}+
  z^{-1}\tau_{-n}^{(1)} q_2^{n}q_1^{-n}\theta \right)  \otimes
  (-)^m z^{-m}E^{-(1)}(z)\left(\tau_{-m}^{(0)} q_2^{m}q_1^{-1-m}+
  z\tau_{-m}^{(1)} q_2^{m}q_1^{-m-2}\theta \right) 
  \]
  \[
   +\res_z  z^{n}E^{+(2)}(z)\left(\tau_{-n}^{(0)} q_2^{n+1}q_1^{-n}+
  \tau_{-n}^{(1)} q_2^{n+1}q_1^{-1-n}\theta \right)  \otimes
   z^{m}E^{-(2)}(z)\left(\tau_{-m}^{(0)} q_2^{m-1}q_1^{-m}+
  \tau_{-m}^{(1)} q_2^{m-1}q_1^{-m-1}\theta \right) 
  \]
  This decomposes in a number of equations of which one is:
  \be
\label{2KP}
  \res_z   \left( (-z)^{n-m}E^{+(1)}(z)\tau_{n}^{(a)}  \otimes
  E^{-(1)}(z)\tau_{m}^{(a)} +
   z^{m-n-2}E^{+(2)}(z)\tau_{n+1}^{(a)}  \otimes
  E^{-(2)}(z)\tau_{m-1}^{(a)} \right) =0
  \ee
  for $a=0,1$ which is a version of the two-component KP equation \cite{JM},\cite{KvdL}, see also \cite{UT}
and
 \be
\label{1mKP}
  \res_z   \left(z^{1-2\delta_{a1}}(-z)^{n-m}E^{+(1)}(z)\tau_{n}^{(a)}  \otimes
  E^{-(1)}(z)\tau_{m}^{(1-a)} +
   z^{m-n-2}E^{+(2)}(z)\tau_{n+1}^{(a)}  \otimes
  E^{-(2)}(z)\tau_{m-1}^{(1-a)} \right) =
  \ee
\[
=\delta_{a1}\tau_n^{(0)}\otimes\tau_{m}^{(1)}
\]
which is a version of the two-component 1st modified KP equation \cite{JM}, \cite{KvdLbispec}.

  We have
  \be\label{tau-0-n}
  \tau_{n}^{(0)} =
  \l n,-n,0| e^{\sum_{n=1}^\infty \left(t_n^{(1)}\alpha_n^{(1)}+t_n^{(2)}\alpha_n^{(2)} \right)}
  g |0,0,0\r\qquad\mbox{and}
  \ee
   \be\label{tau-1-n}
  \tau_{n}^{(1)} =
  \l n-1,-n,1| e^{\sum_{n=1}^\infty \left(t_n^{(1)}\alpha_n^{(1)}+t_n^{(2)}\alpha_n^{(2)} \right)}
  g |0,0,0\r
  \ee

Note, using (\ref{developement}), we can write $g=g^{(0,0)}+g^{(1)}\psi+g^{(-1)}\psi^\dagger +g^{(1,-1)}\psi\psi^\dagger$. Then (\ref{tau-0-n}) and (\ref{tau-1-n}) 
can be written as follows
\be\label{tau-0-nn}
  \tau_{n}^{(0)} =
  \l n,-n,0| e^{\sum_{n=1}^\infty \left(t_n^{(1)}\alpha_n^{(1)}+t_n^{(2)}\alpha_n^{(2)} \right)}
  g^{(0,0)}|0,0,0\r\qquad\mbox{and}
  \ee
   \be\label{tau-1-nn}
  \tau_{n}^{(1)} =
  \l n-1,-n,1| e^{\sum_{n=1}^\infty \left(t_n^{(1)}\alpha_n^{(1)}+t_n^{(2)}\alpha_n^{(2)} \right)}
  g^{(1)} |0,0,0\r
  \ee

 \section{$B$-case in this new realization \label{B-case-in}}
 Recall the involution $\omega$ (see (\ref{omega})) on the Clifford algebra. 
Using the relabeling (\ref{relabel-1})-(\ref{above}),
  this induces
  \be
 \omega(\psi) = \psi^\dag\,,\quad \omega(\psi^\dag) = \psi
  \ee
  \be
  \omega(\psi^{(1)}_{n})=\psi^{\dag(2)}_{-1-n}\,,\quad 
  \omega(\psi^{(2)}_{n})=\psi^{\dag(1)}_{-1-n}\, ,
  \ee
  \be
  \omega(\psi^{\dag(1)}_{n})=\psi^{(2)}_{-1-n}\,,\quad 
  \omega(\psi^{\dag(2)}_{n})=\psi^{(1)}_{-1-n}\, .
  \ee
  It is straightforward to check that
  \be
  \omega(\alpha^{(1)}_n) = -\alpha^{(2)}_n\,,\quad
  \omega(\alpha^{(2)}_n) = -\alpha^{(1)}_n\, .
  \ee
  
  Now define a new basis in our Clifford algebra 
  consisting of elements that are fixed by 
  $\omega$ and of elements $x$ with $\omega(x)=-x$
  
  \be\label{varphi}
  \varphi =\frac{\psi+\psi^\dag}{\sqrt{2}} \,,\qquad {\hat \varphi} =i\frac{\psi-\psi^\dag}{\sqrt{2}} 
  \ee
  \be\label{psi}
  \psi_j =\frac{\psi_j^{(1)}+\psi_{-1-j}^{\dag(2)}}{\sqrt{2}} \,,\qquad 
  {\hat \psi}_j =i\frac{\psi_j^{(1)}-\psi_{-1-j}^{\dag(2)}}{\sqrt{2}}
  \ee
  \be\label{psi-dag}
  \psi_j^\dag =\frac{\psi_j^{\dag(1)}+\psi_{-1-j}^{(2)}}{\sqrt{2}} \,,\qquad 
  {\hat \psi}_j^\dag =-i\frac{\psi_j^{\dag(1)}-\psi_{-1-j}^{(2)}}{\sqrt{2}}
  \ee
  
  Then
   \be
    [\psi_n,\psi_m^\dag]_+=\delta_{n,m}\,,
   \quad [{\hat\psi}_n,{\hat\psi}_m^\dag]_+=\delta_{n,m}
   \ee
   \be
   {\varphi}^2={\hat \varphi}^2=\frac 12
   \ee
 all other elements anticommute. 

 Also
  \be
  \psi_j|0\r = {\hat \psi}_j|0\r =\psi_{-1-j}^\dag|0\r ={\hat \psi}_{-1-j}^\dag|0\r =0\,,
  \quad j<0,
   \ee
   \be
  \l 0|\psi_j^\dag= \l 0|{\hat \psi}_j^\dag =\l 0|\psi_{-1-j} =\l 0|{\hat \psi}_{-1-j}=0\,,
  \quad j<0,
  \ee
   \be
  (\varphi +i{\hat\varphi})|0\r =0 = \l 0|(\varphi - i{\hat\varphi})
  \ee

 \paragraph{Spin module for $b_\infty$.} 
Here we consider the elements $\psi_j,\psi_j^\dag,
 \varphi$,  which are the elements invariant under $\omega$. Recall
  \be
 [\psi_j,\psi_n^\dag]_+=\delta_{j,n}\,,\quad \varphi^2=\frac 12 
  \ee
  all other elements anticommute and
   \be
   \psi_j|0\r =\psi_{1-j}^\dag |0\r =0 =\l 0|\psi_j^\dag = \l 0|\psi_{1-j}\,,\quad j<0
   \ee
 
  The $b_\infty$ spin module splits into two parts $V_{\bar 0} \oplus V_{\bar 1}$ 
  where $V_{\bar 0}$ 
  has the highest weight vector $|{\bar 0}\r = |0\r$, and $V_{\bar 1}$ 
  has the highest weight vector $|{\bar 1}\r = \sqrt{2} \varphi |0\r$.  Both  
  $V_{\bar 0}$ and  $V_{\bar 1}$ are irriducible highest weight modules for $b_\infty$, which is formed by the 
  quadratic elemnts of the   Clifford algebra together with 1.
  
  From now we shall focus on $V_{\bar 0}$.
  
  As before one can define an oscillator algebra $B$
  \be
  \beta_k = \sum_{l\in \mathbb{Z}} :\psi_l\psi_{l+k}^\dag : \, .
  \ee 
  In $V_{\bar 0}$ one has the following highest weight vectors if one restricts to the $B$  
  \be 
|n\r =
\cases{
\Psi_{2l}|0\r \,\qquad \qquad {\rm if }\quad n=2l  \cr
\sqrt{2} \Psi_{2l-1} \varphi |0\r \, \quad \,  {\rm if }\quad n = 2l-1  }
    \\ 
 \,,\qquad
\l n| =
\cases{
\l 0|\Psi_{2l}^\dag \, \qquad \qquad {\rm if }\quad n=2l  \cr
\l 0|\varphi \Psi_{2l-1}^\dag  \sqrt{2} \, \quad \,  {\rm if }\quad n = 2l-1  }
 \ee 
 \be
 \beta_{k}|n\r = 0 = \l 0|\beta_{-k}\,,\quad k>0,\quad n \in\mathbb{Z}
 \ee
 where similar to (\ref{Psi-a})
 \be \label{Psi}
\Psi_n =
\cases{
\psi_{n-1}\cdots \psi_{0} \, \quad {\rm if }\quad n > 0 \cr
1 \qquad\qquad \, \quad \,  {\rm if }\quad n = 0 \cr
\psi^\dag_{n}\cdots \psi^\dag_{-1} \quad \,  {\rm if }\quad n < 0 }
    \\ 
 \,,\quad
\Psi^\dag_n =
\cases{
\psi^\dag_{0}\cdots \psi^\dag_{n-1} \, \qquad {\rm if }\quad n > 0 \cr
1 \qquad\qquad \, \qquad \,  \,{\rm if }\quad n = 0 \cr
\psi_{-1}\cdots \psi_{n} \quad \quad {\rm if }\quad n < 0 }
 \ee

 Let $h$ be a group element of $B_\infty$. Then we can consider $h|{\bar 0}\r$, clearly 
 such an element decomposes
  \be\label{h-dec}
  h|0\r = \sum_{n\in\mathbb{Z}} h_n|n\r
  \ee
 As for $GL_\infty$ there is a boson-fermion correspondence given by
  \be\label{B-bosonization}
  \sigma_B(\psi(z))=pz^{p\frac{\partial}{\partial p}} B^+(z)\,,\quad
  \sigma_B(\psi^\dag(z))=p^{-1}z^{-p\frac{\partial}{\partial p}} B^-(z)
  \ee
  \be\label{B-plus-minus}
  B^\pm(z)=e^{\pm\sum_{n>0} s_n z^n} e^{\mp \sum_{n>0} \frac 1n \frac{\partial}{\partial s_n} z^{-n}}
  \ee
for the fermionic fields (cf. (\ref{FF}))
\be\label{FF2}
\psi(z)=\sum_{i\in \mathbb{Z}}\psi_i z^i,\qquad 
\psi^{\dagger}(z)=\sum_{i\in \mathbb{Z}}\psi_i^{\dagger} z^{-i-1}\, , 
 \ee
and  (\ref{THETA}) induces
  \be
  \sigma_B(\varphi)=\frac{\theta +\frac{\partial}{\partial \theta}}{\sqrt 2}
  \ee
  Then $\sigma_B\left(h|0\r \right)=\tau^B$. And using (\ref{h-dec}) one finds
   \be
   \tau^B = \sum_{l \in\mathbb{Z}} \left( \tau^B_{2l}p^{2l} + \tau^B_{2l-1}p^{2l-1}\theta \right)
   \ee
   Note that $p\theta =-\theta p$.
   
   An element $h\in B_\infty$ is an even element in the Clifford algebra.
  
  We have
  \be
  [h\otimes h, S_B]_-=0\,,\quad 
  S_B =\varphi\otimes\varphi +
  \sum_{i\in\mathbb{Z}} \left(\psi_i\otimes\psi_i^\dag + \psi_i^\dag\otimes\psi_i \right)
  \ee
  Since
  \be
  S_B |0\r\otimes|0\r = \varphi|0\r\otimes\varphi|0\r
  \ee
  One obtains
  \be
  S_B\left(h|0\r\otimes h|0\r  \right)= h\varphi |0\r\otimes h\varphi |0\r
  \ee
 Note that $h=a+\sqrt{2} b\varphi$ where $a$ is an even element expressed in $\psi_i$ and 
 $\psi_i^\dag$, and  $b$ is an odd element expressed in $\psi_i$ and 
 $\psi_i^\dag$
 Then
  \be
\label{BKP}
  S_B\left(a+\sqrt{2} b\varphi  \right)|0\r\otimes\left(a+\sqrt{2} b\varphi  \right)|0\r=
  \left(a\varphi +\frac 12\sqrt{2} b  \right)|0\r\otimes
  \left(a\varphi+\frac 12\sqrt{2} b  \right)|0\r
  \ee
 or
  \[
  \left(a\varphi -\frac 12\sqrt{2} b  \right)|0\r\otimes\left(a\varphi -
  \frac 12\sqrt{2} b  \right)|0\r +
  \]
  \[
 \sum_{i\in\mathbb{Z}} \psi_i  \left(a +\sqrt{2} b\varphi  \right)|0\r\otimes
 \psi_i^\dag\left(a+\sqrt{2} b\varphi  \right)|0\r +
  \]
   \[
 \sum_{i\in\mathbb{Z}} \psi_i^\dag  \left(a+\sqrt{2} b\varphi  \right)|0\r\otimes
 \psi_i\left(a+\sqrt{2} b\varphi  \right)|0\r
  \]
  \[
  =\left(a\varphi +\frac 12\sqrt{2} b  \right)|0\r\otimes\left(a\varphi +
  \frac 12\sqrt{2} b  \right)|0\r 
  \]

  Hence
  \[
  \sum_{i\in\mathbb{Z}} \left[ \psi_i(a+\sqrt{2} b\varphi)|0\r\otimes
  \psi_i^\dag(a+\sqrt{2}b\varphi)|0\r  +
  \psi_i^\dag(a+\sqrt{2} b\varphi)|0\r\otimes
  \psi_i(a+\sqrt{2}b\varphi)|0\r  \right]
  \]
  \[
  =\sqrt{2} a \varphi |0\r\otimes b|0\r + b|0\r\otimes\sqrt{2} a\varphi |0\r
  \]
  Now clearly:
  \be
  \sigma\left( a|0\r \right) =\sum_{l\in\mathbb{Z}} \tau^B_{2l} p^{2l}\,,\quad
   \sigma\left(\sqrt{2} b\varphi |0\r \right) =\sum_{l\in\mathbb{Z}} \tau^B_{2l-1} p^{2l-1}\theta
  \ee
  \be
  \sigma\left( \sqrt{2} a\varphi |0\r \right) =\sum_{l\in\mathbb{Z}} \tau^B_{2l} p^{2l}\theta\,,\quad
   \sigma\left(b |0\r \right) =\sum_{l\in\mathbb{Z}} \tau^B_{2l-1} p^{2l-1}
  \ee
Also
 \be
 \tau^B_l(s) = \l l|e^{\sum_{l}s_l\beta_l} h |0\r
 \ee
 
Equation (\ref{BKP}) turns into the large BKP hierarchy or Pfaff lattice:
\begin{equation}
\label{BKP2}
 \res_z  
\left( 
z^{n-m-2}B^+(z)\tau_{n-1}^B\otimes
B^-(z)\tau_{m+1}^B
+\frac{(-1)^{n+m}}{2z} 
\tau_{n}^B\otimes
\tau_{m}^B+z^{m-n-2}B^-(z)\tau_{n+1}^B\otimes
B^+(z)\tau_{m-1}^B
\right)
=\frac12\tau_{n}^B\otimes
\tau_{m}^B
\end{equation}

\section{A relation between the tau functions}

In the same way as the small BKP is related to  the KP hierarchy, the large BKP (or Pfaff lattice) is 
related to the 2-component KP (or Toda Lattice hierarchy). In fact one can use the same involution $\hat{}$
 of (\ref{hat}) on the
 Clifford algebra, it induces
  \be
  \hat{}\, (\varphi)={\hat \varphi}\,,\quad \hat{}\, ({\hat \varphi})=\varphi
  \ee
  \be
  \hat{}\, (\psi_i)={\hat \psi}_i\,,\quad \hat{}\, (\hat\psi_i)={\psi}_i\,,\quad 
  \hat{}\, (\psi_i^\dag)={\hat \psi}_i^\dag\,,\quad \hat{}\, ({\hat\psi}_i^\dag)={\hat \psi}_i^\dag
  \ee
  Then
  \be
  \hat{}\, (\beta_l)={\hat\beta}_l=\sum_{j\in\mathbb{Z}} :{\hat\psi}_j{\hat\psi}_{j+l}^\dag :
  \ee
 We want to consider
  \be
  e^{\sum_{l} s_l(\beta_l+\hat{}\, (\beta_l))} h\cdot \hat{}\,(h)\,|0\r=
  e^{\sum_{l} s_l\beta_l} h\cdot e^{\sum_{l} s_l{\hat \beta}_l}{\hat h}\,|0\r
  \ee
 Now
  \be
  \psi_j\psi_k^\dag + {\hat \psi}_j{\hat \psi}_k^\dag=\frac 12
 \left(\psi_j^{(1)}+ \psi_{-j-1}^{\dag(2)}\right) 
 \left(\psi_k^{\dag(1)}+ \psi_{-k-1}^{(2)}\right) 
+  \frac 12\left(\psi_j^{(1)}- \psi_{-j-1}^{\dag(2)}\right) 
\left(\psi_k^{\dag(1)}- \psi_{-k-1}^{(2)}\right) 
  \ee
\[
 =\psi_j^{(1)}\psi_k^{\dag(1)}+\psi_{-j-1}^{(2)}\psi_{-k-1}^{\dag(2)}
\]
Thus
\[
 \beta_l+{\hat \beta}_l =\sum_j \left(:\psi_j\psi_{j+l}^\dag : + 
 :{\hat\psi}_j{\hat\psi}_{j+l}^\dag :\right)=\sum_j \left(:\psi_j^{(1)}\psi_{j+l}^{\dag(1)}: + 
 :\psi_{-j-1}^{\dag(2)} \psi_{-j-l-1}^{(2)}:\right)
\]
\[
 =\alpha_l^{(1)} - \alpha_j^{(2)}
\]
and hence  $e^{\sum s_l(\beta_l+{\hat \beta}_l)}= 
e^{\sum_l s_l\alpha_l^{(1)}}e^{-\sum_l s_l\alpha_l^{(2)}}$

 First note that
  \be
  \psi=\frac{\varphi - i {\hat\varphi}}{\sqrt{2}}\,,\quad
  \psi^\dag=\frac{\varphi + i {\hat\varphi}}{\sqrt{2}}
  \ee
  \be
  \psi_j^{(1)}=\frac{\psi_{j} -i {\hat \psi}_{j}}{\sqrt 2} \,,\quad
  \psi_j^{\dag(1)}=\frac{\psi_{j}^\dag +i {\hat \psi}_{j}^\dag}{\sqrt{2}}
  \ee
  \be
  \psi_j^{(2)}=\frac{\psi_{-j-1}^\dag -i {\hat \psi}_{-j-1}^\dag}{\sqrt{2}} \,,\quad
  \psi_j^{\dag(2)}=\frac{\psi_{-j-1} +i {\hat \psi}_{-j-1}}{\sqrt{2}}
  \ee
Thus
 \be
 \psi_j^{(1)}\psi_{-j-1}^{\dag(2)}=
 \frac 12 \left(\psi_j-i{\hat \psi}_j \right)\left(\psi_j+i{\hat \psi}_j \right)
 =i\psi_j{\hat \psi}_j
 \ee
 \be
 \psi_j^{\dag(1)}\psi_{-j-1}^{(2)}=
 \frac 12 \left(\psi_j^\dag +i{\hat \psi}_j^\dag \right)
 \left(\psi_j^\dag -i{\hat \psi}_j^\dag \right)
 =-i\psi_j^\dag{\hat \psi}_j^\dag
 \ee
The following lemma will be useful:
  \begin{Lemma}\label{lemma-vac=vac}
 \be\label{vac=vac-even}
   \l -2n,2n,0|=(-1)^n \l 0|\Psi_{-2n}^\dag {\hat \Psi_{-2n}^\dag }
   \ee
 \be\label{vac=vac-odd}
  \l 1-2n,2n-1,0|=(-1)^{n+1}\l 0|\left(\sqrt{2}\varphi\Psi_{1-2n}^\dag  \right)
  \left(\sqrt{2}{\hat \varphi}{\hat \Psi}_{1-2n}^\dag  \right)
 \ee
\end{Lemma}
\noindent {\bf Proof.} 
Let $n>0$. Then
  \[
   \l -2n,2n,0|=\l 0|\Psi_{-2n}^{\dag(1)}\Psi_{2n}^{\dag(2)}=
   \l 0|\psi_{-1}^{(1)}\psi_{-2}^{(1)}\cdots \psi_{-2n}^{(1)}\psi_0^{\dag(2)}\psi_1^{\dag(2)}
   \cdots \psi_{2n-1}^{\dag(2)}
  \]
  \[
   =(-1)^n \l 0|\psi_{-1}\cdots \psi_{-2n}{\hat \psi}_{-1}\cdots {\hat \psi}_{-2n}=
   (-1)^n \l 0|\Psi_{-2n}^\dag {\hat \Psi_{-2n}^\dag }
  \]
If $n<0$ then we have
 \[
   \l -2n,2n,0|=\l 0|\Psi_{-2n}^{\dag(1)}\Psi_{2n}^{\dag(2)}=
   \l 0|\psi_{0}^{\dag(1)}\psi_{1}^{\dag(1)}\cdots \psi_{-2n-1}^{\dag(1)}
   \psi_{-1}^{(2)}\psi_{-2}^{(2)}   \cdots \psi_{2n}^{(2)}
  \]
  \[
   =(-1)^n \l 0|\psi_{0}^\dag\psi_{1}^\dag\cdots \psi_{-2n-1}^\dag
   {\hat \psi}_{0}^\dag\cdots {\hat \psi}_{-2n-1}^\dag=
   (-1)^n \l 0|\Psi_{-2n}^\dag {\hat \Psi_{-2n}^\dag }
  \]
   Next we consider the case
   \be
   \l 1-2n,2n-1,0| =\l 0|\Psi_{-2n+1}^{\dag(1)}\Psi_{2n-1}^{\dag(2)}
   \ee
   If $n>0$ then:
   \[
   \l 1 -2n,2n-1,0|=
   \l 0|\psi_{-1}^{(1)}\cdots \psi_{-2n+1}^{(1)}
   \psi_{0}^{\dag(2)}   \cdots \psi_{2n-2}^{\dag(2)}
  \]
  \[
   =(i)^{2n-1} \l 0|\psi_{-1}\cdots \psi_{-2n+1}
   {\hat \psi}_{-1}\cdots {\hat \psi}_{-2n+1}=
   (i)^{2n-1} \l 0|\Psi_{-2n+1}^\dag {\hat \Psi_{-2n+1}^\dag }
  \]
    Now use that
  \be
  \l 0|\varphi{\hat \varphi} =\frac i2 \l 0|(\psi+\psi^\dag)(\psi+\psi^\dag) =
  -\frac i2\l 0|\psi^\dag\psi=-\frac i2 \l 0|
  \ee
 Thus
 \[ 
  \l1 -2n,2n-1,0|= 
  2(i)^{2n} \l 0|\varphi{\hat \varphi}\Psi_{-2n+1}^\dag {\hat \Psi_{-2n+1}^\dag }
  =(-1)^{n+1}\l 0|\left(\sqrt{2}\varphi\Psi_{1-2n}^\dag  \right)
  \left(\sqrt{2}{\hat \varphi}{\hat \Psi}_{1-2n}^\dag  \right)
 \]
  If $n<0$ then
  \[
   \l 1-2n,2n-1,0|=
   \l 0|\psi_{0}^{\dag(1)}\cdots \psi_{2n}^{\dag(1)}
   \psi_{-1}^{(2)}   \cdots \psi_{2n-1}^{(2)}
  \]
  \[
   =(-i)^{1-2n} \l 0|\psi_{0}^\dag\cdots \psi_{-2n}^\dag
   {\hat \psi}_{0}^\dag\cdots {\hat \psi}_{-2n}^\dag=
   (-i)^{1-2n} \l 0|\Psi_{1-2n}^\dag {\hat \Psi_{1-2n}^\dag }
  \]
  \[
=(-1)^{n+1}\l 0|\left(\sqrt{2}\varphi\Psi_{1-2n}^\dag  \right)
  \left(\sqrt{2}{\hat \varphi}{\hat \Psi}_{1-2n}^\dag  \right)
 \]
Which finishes the proof of the lemma \hfill$\square$
\ \\

   Thus
   \be
   \tau_{-2n}^{(0)}(s,-s)=(-1)^n \l 0|\Psi_{-2n}^\dag {\hat \Psi_{-2n}^\dag }
   e^{\sum_k s_k\beta_k}e^{\sum_k s_k{\hat \beta}_k} h{\hat h}|0\r
   \ee
      \be
  =(-1)^n 
  \l -2n| e^{\sum_k s_k\beta_k} h|0\r   
  \l {\widehat {-2n}}| e^{\sum_k s_k{\hat \beta}_k} {\hat h}|0\r=
  (-1)^n \tau_{2n}^B(s)\tau_{2n}^B(s)
   \ee
And 
 \[
   \tau_{1-2n}^{(0)}(s,-s)=
   \l 1-2n,2n-1,0|e^{\sum_k s_k\beta_k}e^{\sum_k s_k{\hat \beta}_k} h{\hat h}|0\r =
 \]
   \[
  =(-1)^{n+1}\l 1-2n| e^{\sum_k s_k\beta_k} h|0\r   
  \l {\widehat {1-2n}}| e^{\sum_k s_k{\hat \beta}_k} {\hat h}|0\r
  =(-1)^{n+1}\tau_{1-2n}^B(s)\tau_{1-2n}^B(s)
   \]
  Thus
\begin{Proposition} \label{Pfaff-det-tau}
A BKP tau function satisfies
\[
 \left(\tau_{n}^B(s)\right)^2=(-)^{\frac{n(n+1)}{2}} \tau_{n}^{(0)}(s,-s)
\]
\end{Proposition}

  \br\label{g1g2=h1h2}
  If $g_i=h_i{\hat h}_i$, $i=1,\dots,k$,  then $g= g_1 \cdots g_k = h{\hat h}$ where $h= h_1\cdots h_k$ and 
  ${\hat h}={\hat h}_1 \cdots {\hat h}_k$.
  \er
\br 
It follows from Proposition \ref{Pfaff-det-tau}  that the square of the BKP  function
 \be\label{BKP-wave-function}
 V_n(t,z):=e^{\sum_{m=1}^\infty t_m z^m}\frac{\tau_{n+1}\left(t-[z^{-1}]\right)}{\tau_n(t)}
 \ee
 is related to the two-component KP Green function
  \be
  K_n(x,y,t^{(1)},t^{(2)}):=
  \l n+1,n-1|e^{\sum_{m>1} \left( t_m^{(1)}\alpha_m^{(1)}+t_m^{(2)}\alpha_m^{(2)}\right)}\psi^{(1)}(x)
  \psi^{\dag(2)}(y) h{\hat h}|0,0\r
  \ee
  as follows
  \be\label{BKPwave-function-square}
  \left(V_n(t,z)\right)^2\,=\,K_n(z,z,t,-t)
  \ee
  Let us note that two-component KP is useful to study matrix models. The Green function $K(x,y)$ is widely
  used for computing various correlation functions. However, these  models  do not possess the property
  of factorization $g=h{\hat h}$.
\er

  \section{A relation between the wave functions}

We introduce the 2 component KP wave function $W_n^{\pm (a)}(t,z)$ by
\[
W_n^{\pm (a)}(t,z)=\frac1{\tau_{n}^{(a)}(t)}\left(
\begin{array}{cc}
(-z)^{\pm n}E^{\pm(1)}(z)\tau_{n}^{(a)}(t) &
 z^{\mp n-1}E^{\pm(2)}(z)\tau_{n\pm 1}^{(a)}(t) \\[2mm]
(-z)^{\pm n-1}E^{\pm(1)}(z)\tau_{n\mp 1}^{(a)}(t) &
 z^{\mp n}E^{\pm(2)}(z)\tau_{n}^{(a)}(t)
\end{array}
\right)
\]
then (\ref{2KP}) turns into
\be
\label{2KP2}
 \res_z  W_n^{+(a)}(s,z)W_m^{-(a)}(t,z)^T
=0
\ee
and (\ref{1mKP}) turns into
\be
\label{2KP2-}
 \res_z  W_n^{+(a)}(s,z)
\left(
\begin{array}{cc}
z^{1-2\delta_{a1}}
&
0\\
0&1
\end{array}
\right)
W_m^{-(1-a)}(t,z)^T
=\frac{\delta_{a1}}{\tau_n^{(1)}(s)\tau_m^{(0)}(t)}\left(
\begin{array}{c}
\tau_n^{(0)}(s)\\
\tau_{n-1}^{(0)}(s)
\end{array}\right)
\left(
\begin{array}{cc}
\tau_{m}^{(1)}(t)&
\tau_{m+1}^{(1)}(t)

\end{array}
\right)
\ee
We introduce next the BKP wave function $V_n^\pm(z)$. To do that we   first 
observe that (\ref{BKP2}) can be rewritten in the matrix form
\begin{equation}
\label{BKP3} 
\res_z 
R^+_n(z)S(z)R^-_m(z)^T=\res_z T^+_nS(z)T_m^{-T}
\end{equation}
where
\begin{equation}
R_n^\pm (z)=\left(
\begin{array}{ccc}
z^{\pm n}B^\pm(z)\tau_{n}^B & 
(-1)^{n+1}\tau_{n\pm 1}^B & 
z^{\mp n-2}B^\mp (z)\tau_{n\pm 2}^B \\
z^{\pm n-1}B^\pm(z)\tau_{n\mp 1}^B & 
(-1)^{n}\tau_{n}^B & 
z^{\mp n-1}B^\mp(z)\tau_{n\pm 1}^B \\
z^{\pm n-2}B^\pm (z)\tau_{n\mp 2}^B & 
(-1)^{n-1}\tau_{n\mp 1}^B & 
z^{\mp n}B^\mp (z)\tau_{n}^B 
\end{array}
\right),
\quad
T_n^\pm=\left(\begin{array}{ccc}
\tau_{n}^B & 
\tau_{n\pm 1}^B & 
0\\
0 & 
\tau_{n}^B & 
0 \\
0 & 
\tau_{n\mp 1}^B & 
\tau_{n}^B 
\end{array}
\right)
\end{equation}
and
\[
S(z)=\mbox{diag}(1,\frac1{2z}, 1)
\]
Let 
\[
{U^\pm_n}=
\left(\begin{array}{ccc}
\frac1{\tau_{n}^B} & 
\mp\frac{\tau_{n\pm 1}^B}{(\tau_{n}^{B})^2} & 
0\\[2mm]
0 & 
 \frac1{\tau_{n}^B} & 
0 \\[2mm]
0 &\mp\frac{\tau_{n\mp 1}^B }{(\tau_{n}^{B})^2}
 &  \frac1{\tau_{n}^B}
\end{array}
\right)
\]
then 
$U^+_n=({T^+_n})^{-1}$.
Now introduce the  BKP wave function  $V_n^\pm(z)=U^\pm_nR^\pm_n(z)$ then (\ref{BKP3}) turns into 
\be
\label{BKP4}
\res_z V_n^+(z)S(z)V_m^-(z)^T=\left(\begin{array}{ccc}
0&0&0\\
\frac{\tau_{m- 1}^B}{\tau_{m}^B}&\frac12 &\frac{\tau_{m+1}^B}{\tau_{m}^B}\\0&0&0
\end{array}
\right)
\ee
We will now  show that two-component KP wave functions $W_n^{\pm(0)}(t^{(1)},t^{(2)},z)$
evaluated at $t^{(1)}_j=s_j=-t^{(2)}$  for  group elements, $g=h\hat h$ may be expressed in terms of the BKP wave 
functions $V_n^\pm(z)$. 
\begin{Proposition}
\label{Prop2}
A BKP wave function is related to a 2 component KP wave function, via
\[
W_n^{\pm(0)}(s,-s,z)=
\left(
\begin{array}{ccc}
(-1)^n&0&0\\
0&0&\mp 1
\end{array}
\right)
V_n^\pm(s,z)
\left(
\begin{array}{cc}
1&0\\
0&0\\
0&\mp 1
\end{array}
\right)
\]
\end{Proposition}
 To prove this we calculate 
  \be
 w_n^{(1)}(t^{(1)},t^{(2)},z)=\l n+1,-n,0|
 e^{\sum_{j>0}\left(t^{(1)}_j\alpha_j+t^{(2)}_j{\hat\alpha}_j\right)} \psi^{(1)}(z)g |0\r
 \ee
 \be
 =(-z)^ne^{\sum_{j>0}t^{(1)}_jz^j}\l n,-n,0|e^{
 \sum_{j>0}\left(\left(t^{(1)}_j-\frac{z^{-j}}{j} \right)\alpha_j+t^{(2)}_j{\hat\alpha}_j\right)}
g |0\r
 \ee
  \be
 =(-z)^ne^{\sum_{j>0}t^{(1)}_jz^j}\tau_n^{(0)}\left(t^{(1)}-[z^{-1}], t^{(2)}\right)
 \ee
  Now set $g=h{\hat h}$ and $t^{(1)}_j=s_j=-t^{(2)}$. Then 
 \[
 w_n^{(1)}(s,-s,z)=(-z)^ne^{\sum_{j>0}s_jz^j}\tau_n^{(0)}\left(s-[z^{-1}], -s\right)
  \]
  \[
 =\l n+1,-n,0|e^{\sum_{j>0}s_j(\beta_j+{\hat\beta}_j)} \psi^{(1)}(z) h{\hat h}|0\r
 \]
 \[
  =(-i)^{n+1}\l 0|\Psi^\dag_{n+1} {\hat\Psi}^\dag_{n+1}\psi_{-n-1}^{\dag(2)}
  e^{\sum_{j>0}s_j(\beta_j+{\hat\beta}_j)} \psi^{(1)} h{\hat h}|0\r
 \]
 \[
  =\frac{(-i)^{n+1}}{2}\l 0|\Psi^\dag_{n+1} {\hat\Psi}^\dag_{n+1}(\psi_n+i{\hat\psi}_n)
  e^{\sum_{j>0}s_j(\beta_j+{\hat\beta}_j)} (\psi(z)-i{\hat\psi}(z)) h{\hat h}|0\r
 \]
   \[
  =\left(\frac{i^{n+1}}{2}\l 0|\Psi^\dag_{n} {\hat\Psi}^\dag_{n+1}+
  \frac{(-i)^{n}}{2}\l 0|\Psi^\dag_{n} {\hat\Psi}^\dag_{n+1}
  \right)
  e^{\sum_{j>0}s_j(\beta_j+{\hat\beta}_j)} (\psi(z)-i{\hat\psi}(z)) h{\hat h}|0\r
   \]
   \[
  =\frac{(-i)^{n+1}}{2} z^{n-1} e^{\sum_{j>0} s_jz^j}\l 0|\Psi^\dag_{n-1} {\hat\Psi}^\dag_{n+1}
  e^{\sum_{j>0}\left(s_j-\frac{z^{-j}}{j}  \right)\beta_j+ s_j{\hat\beta}_j)}  h{\hat h}|0\r
   \]
    \[
  +\frac{i^{n}}{2} z^{n} e^{\sum_{j>0} s_jz^j}\l 0|\Psi^\dag_{n} {\hat\Psi}^\dag_{n}
  e^{\sum_{j>0}\left(s_j-\frac{z^{-j}}{j}\right)\beta_j +s_j {\hat\beta}_j)}  h{\hat h}|0\r
   \]
    \[
  +\frac{i^{n}}{2} z^{n} e^{\sum_{j>0} s_jz^j}\l 0|\Psi^\dag_{n} {\hat\Psi}^\dag_{n}
  e^{\sum_{j>0} s_j \beta_j +\left(s_j-\frac{z^{-j}}{j}\right){\hat\beta}_j)}  h{\hat h}|0\r
   \]
   \[
  +\frac{(-i)^{n+1}}{2} z^{n-1} e^{\sum_{j>0} s_jz^j}\l 0|\Psi^\dag_{n+1} {\hat\Psi}^\dag_{n-1}
  e^{\sum_{j>0} s_j \beta_j +\left(s_j-\frac{z^{-j}}{j}\right){\hat\beta}_j)}  h{\hat h}|0\r
   \] 
   \be
   =\cases{ (-1)^{\frac n2}z^n e^{\sum_{j>0} s_jz^j}
   \left(\tau^B_n(s)\tau^B_n(s-[z^{-1}]) -z^{-1}\tau^B_{n+1}(s)\tau^B_{n-1}(s-[z^{-1}])  \right)
   \,,\quad {\rm if}\quad n\,{\rm even}
   \cr
   (-1)^{\frac {n-1}{2}}z^n e^{\sum_{j>0} s_jz^j}
   \left(\tau^B_n(s)\tau^B_n(s-[z^{-1}]) -z^{-1}\tau^B_{n+1}(s)\tau^B_{n-1}(s-[z^{-1}])  \right)
  \,,\quad {\rm if}\quad n\,{\rm odd}
  }
   \ee
   Thus we obtain
   \be\label{}
  w_n^{(1)}(s,-s,z)=(-1)^{\frac{n(n-1)}{2}}z^n e^{\sum_{j>0} s_jz^j}
   \left(\tau^B_n(s)\tau^B_n(s-[z^{-1}]) -z^{-1}\tau^B_{n+1}(s)\tau^B_{n-1}(s-[z^{-1}])  \right) 
   \ee
 which is the first formula of
   \be
\label{xxx}
 (-1)^{\frac{n(n+1)}{2}}\tau_n^{(0)}\left(s-[z^{-1}], -s\right) = 
 \tau^B_n(s)\tau^B_n(s-[z^{-1}]) -z^{-1}\tau^B_{n+1}(s)\tau^B_{n-1}(s-[z^{-1}]) 
   \ee   
   \be
  (-1)^{\frac{n(n+1)}{2}}\tau_n^{(0)}\left(s+[z^{-1}], -s\right) =
  \tau^B_n(s)\tau^B_n(s+[z^{-1}]) + z^{-1}\tau^B_{n-1}(s)\tau^B_{n+1}(s+[z^{-1}]) 
   \ee
   \be
 (-1)^{\frac{n(n+1)}{2}} \tau_n^{(0)}\left(s, -s-[z^{-1}]\right) =
 \tau^B_n(s)\tau^B_n(s+[z^{-1}]) -z^{-1}\tau^B_{n-1}(s)\tau^B_{n+1}(s+[z^{-1}]) 
   \ee
   \be
\label{xxxx}
  (-1)^{\frac{n(n+1)}{2}}\tau_n^{(0)}(s,-s + [z^{-1}])) =
  \tau^B_n(s)\tau^B_n(s-[z^{-1}]) + z^{-1}\tau^B_{n+1}(s)\tau^B_{n-1}(s-[z^{-1}])
   \ee
The other formulas can be obtained in a similar way. From these formulas and Proposition \ref{Pfaff-det-tau}
one easily deduces Proposition \ref{Prop2}.

   \section{The two-sided BKP (2-BKP) and two-component Toda lattice}
In this section we consider a two sided version of some of the previous constructions.
   
 \paragraph{2-BKP and two-component 2-KP (two-component Toda lattice) tau functions.}

  Consider also
  \be\label{two-component-TL-tau}
  \tau_{n,m,l}^{(0)}(t^{(1)},t^{(2)};{\bar t}^{(1)},{\bar t}^{(2)}|g):= 
  \l n,l-n,0 |e^{\sum_{i>0} (t^{(1)}_i\alpha^{(1)}_i + t^{(2)}_i\alpha^{(2)}_i)}\, g \,
  e^{ \sum_{i>0} ({\bar t}^{(1)}_i\alpha^{(1)}_{-i} + {\bar t}^{(2)}_i\alpha^{(2)}_{-i})  } |m,l-m,0 \r
  \ee
  which may be considered as the two-component 2-KP tau function, or, the same, two-component Toda lattice tau 
  function\footnote{see a piece 
  between relations (9.6) and (9.7) in \cite{JM} }. The Hirota equations for the tau function (\ref{two-component-TL-tau})
  may  also be found in the Appendix \ref{two-sided-Hirota}.
   \br\label{two-component-TL-involution} We have
  \be
  \tau_{n,m,l}^{(0)}(t^{(1)},t^{(2)};{\bar t}^{(1)},{\bar t}^{(2)}|g)= 
   \tau_{m,n,l}^{(0)}({\bar t}^{(1)},{\bar t}^{(2)};t^{(1)},t^{(2)}|g^\dag)
  \ee
   \er
   For the proof of the Remark \ref{two-component-TL-involution} we notice that
   \[
    \left(\l n,l-n,0 |e^{\sum_{i>0} (t^{(1)}_i\alpha^{(1)}_i + t^{(2)}_i\alpha^{(2)}_i)} \right)^\dag =
    e^{\sum_{i>0} (t^{(1)}_i\alpha^{(1)}_{-i} + t^{(2)}_i\alpha^{(2)}_{-i})}|n,l-n,0\r
   \]
   \[
    \left( g \,
  e^{ \sum_{i>0} ({\bar t}^{(1)}_i\alpha^{(1)}_{-i} + {\bar t}^{(2)}_i\alpha^{(2)}_{-i})  } |m,l-m,0 \r\right)^\dag = 
  \l m,l-m,0  | e^{ \sum_{i>0} ({\bar t}^{(1)}_i\alpha^{(1)}_{i} + {\bar t}^{(2)}_i\alpha^{(2)}_{i})  } \,g^\dag 
   \]
and the fact that the pairing of two vectors and of two corresponding dual vectors coincides.
   
   Later we shall omit the argument $g$ on the left hand side of (\ref{two-component-TL-tau}).

   In what follows we shall 
  put $l=0$,
  ${ t}^{(1)}_k={ s}_k=-{ t}^{(2)}_k$, 
  ${\bar t}^{(1)}_k={\bar s}_k=-{\bar t}^{(2)}_k$, where $s$ and ${\bar s}$ are two independent sets of variables.
  Below $g=h{\hat h}$. These are restrictions which allow to compare two-component TL tau functions with BKP tau 
  functions.

   Consider a 2-BKP tau function
  \be\label{2BKP-tau}
  \tau^B_{n,m}(s,{\bar s}|h):= {'\l} n |e^{\sum_{i>0} s_i\beta_i} \,h\, e^{\sum_{i>0} {\bar s}_i\beta_{-i}} |m{\r}'
  \ee
  which depends on two discrete parameters $n$ and $m$ and two sets of higher times $s=(s_1,s_2,\dots)$ and 
  ${\bar s}=({\bar s}_1,{\bar s}_2,\dots)$. Hirota equations of the 2-BKP. The name 2-BKP (two-sided BKP) is related to 
  the fact that 
  2-BKP Hirota equations (see the Appendix \ref{two-sided BKP-Hirota}) contains the BKP Hirota
  equations with respect to the variables $s,m$ the same as Hirota BKP equations with respect the variables ${\bar s},n$
   (\ref{BKP2}).

   \br\label{2BKP-involution} We have the following symmetry
  \be
  \tau_{n,m}^B(s,{\bar s}|h)= 
   \tau_{m,n}^B({\bar s},s|h^\dag)
  \ee
   which is proved in the same way as the Remark \ref{two-component-TL-involution}.       
   Then it follows that the 2-BKP tau functions (\ref{2BKP-tau}) $\tau_{n,m}(s,{\bar s})$ is a BKP tau 
   function with respect to the variables ${\bar s}$. This explains the name 2-BKP tau function.
  \er

  In the same way as  Proposition \ref{Pfaff-det-tau} was obtained we get
  
  \begin{Proposition} \label{Pfaff-det-2tau}
  A BKP tau function satisfies
\be\label{Pfaff-det-2-tau}
 \left( \,\tau_{n,m}^B(s,{\bar s}|h)\,\right)^2=
 (-)^{\frac{n(n+1)}{2}+\frac{m(m+1)}{2}} \tau_{n,m,0}^{(0)}(s,-s;{\bar s},-{\bar s}|g)
\ee
where $g=h{\hat h}$.

 \end{Proposition}

For proof of Proposition \ref{Pfaff-det-2tau} we note that from Lemma \ref{lemma-vac=vac} it follows that
   \[
    |-2n,2n,0\r = (-1)^n {\hat \Psi}_{-2n} \Psi_{-2n} |0\r\,,\qquad 
     |1-2n,2n-1,0\r = (-1)^{n+1} {\hat \Psi}_{1-2n}\sqrt{2}{\hat\varphi} \Psi_{1-2n}\sqrt{2}{\varphi} |0\r
   \]
which allows to repeat all the steps of the derivation of the Proposition \ref{Pfaff-det-tau}.
   
The other way to prove it is just to modify $g$ in the Proposition \ref{Pfaff-det-tau} by a certain right factor
whose action on $|0,0,0\r$  recreates the vector $(-1)^{\frac{m(m+1)}{2}}|-m,m,0\r$.

   \paragraph{Miwa transforms}
   
   For tau functions of the two-component TL and 2-BKP we can write (cf. (\ref{xxx}-\ref{xxxx})):
    \be
\label{xxx2BKP-left-1}
 (-1)^{\frac{n(n+1)}{2}+\frac{m(m+1)}{2}}\tau_{n,m,0}^{(0)}\left(s-[x], -s; {\bar s},-{\bar s}\right) 
 = \tau^B_{n,m}(s, {\bar s})\tau^B_{n,m}(s-[x], {\bar s}) 
 -x\tau^B_{n+1,m}(s, {\bar s})\tau^B_{n-1,m}(s-[x], {\bar s}) 
   \ee   
   \be\label{xxx2BKP-left-2}
  (-1)^{\frac{n(n+1)}{2}+\frac{m(m+1)}{2}}\tau_{n,m,0}^{(0)}\left(s+[x], -s; {\bar s},-{\bar s}\right) 
  =\tau^B_{n,m}(s, {\bar s})\tau^B_{n,m}(s+[x], {\bar s}]) 
  + x\tau^B_{n-1,m}(s, {\bar s})\tau^B_{n+1, m}(s+[x], {\bar s}) 
   \ee
   \be\label{xxx2BKP-left-3}
 (-1)^{\frac{n(n+1)}{2}+\frac{m(m+1)}{2}} \tau_{n,m,0}^{(0)}\left(s, -s-[x]; {\bar s},-{\bar s}\right) 
 =\tau^B_{n,m}(s, {\bar s})\tau^B_{n,m}(s+[x], {\bar s}) 
 -x\tau^B_{n-1,m}(s, {\bar s})\tau^B_{n+1,m}(s+[x], {\bar s}) 
   \ee
   \be\label{xxx2BKP-left-4}
  (-1)^{\frac{n(n+1)}{2}+\frac{m(m+1)}{2}}\tau_{n,m,0}^{(0)}\left(s,-s + [x]; {\bar s},-{\bar s}\right) 
  =\tau^B_{n,m}(s, {\bar s})\tau^B_{n,m}(s-[x], {\bar s}) 
  + x\tau^B_{n+1,m}(s, {\bar s})\tau^B_{n-1,m}(s-[x], {\bar s})
   \ee
  Let us note that it is correct for any choice of $h$. Then after replacing $g=h{\hat h}$ by 
  $g^\dag={\hat h}^\dag h^\dag= h^\dag{\hat h}^\dag $
  in the last formulae and using the Remarks \ref{two-component-TL-involution}, \ref{2BKP-involution} from 
  (\ref{xxx2BKP-left-1})-(\ref{xxx2BKP-left-4}) it may be obtained:
    \be\label{xxx2BKP-right-1}
   (-1)^{\frac{n(n+1)}{2}+\frac{m(m+1)}{2}}\tau_{n,m,0}^{(0)}\left(s, -s; {\bar s}-[x],-{\bar s}\right) 
   = \tau^B_{n,m}(s, {\bar s})\tau^B_{n,m}(s, {\bar s}-[x]) 
   -x\tau^B_{n,m+1}(s, {\bar s})\tau^B_{n,m-1}(s, {\bar s}-[x]) 
   \ee   
   \be\label{xxx2BKP-right-2}
   (-1)^{\frac{n(n+1)}{2}+\frac{m(m+1)}{2}}\tau_{n,m,0}^{(0)}\left(s, -s; {\bar s}+[x],-{\bar s}\right) 
   =\tau^B_{n,m}(s, {\bar s})\tau^B_{n,m}(s, {\bar s}+[x]) 
   + x\tau^B_{n,m-1}(s, {\bar s})\tau^B_{n, m+1}(s, {\bar s}+[x]) 
   \ee
   \be\label{xxx2BKP-right-3}
   (-1)^{\frac{n(n+1)}{2}+\frac{m(m+1)}{2}} \tau_{n,m,0}^{(0)}\left(s, -s; {\bar s},-{\bar s}-[x]\right) 
   =\tau^B_{n,m}(s, {\bar s})\tau^B_{n,m}(s, {\bar s}+[x]) 
   -x\tau^B_{n,m-1}(s, {\bar s})\tau^B_{n,m+1}(s, {\bar s}+[x])  
   \ee
   \be\label{xxx2BKP-right-4}
  (-1)^{\frac{n(n+1)}{2}+\frac{m(m+1)}{2}}\tau_{n,m,0}^{(0)}\left(s,-s); {\bar s},-{\bar s} + [x]\right) 
  =\tau^B_{n,m}(s, {\bar s})\tau^B_{n,m}(s, {\bar s}-[x]) 
  + x\tau^B_{n,m+1}(s, {\bar s})\tau^B_{n,m-1}(s, {\bar s}-[x])
   \ee

    \section{An example: Toda lattice and B-type Pfaff lattice (BPL) }
  
    In many applications (like random matrices or random partitions) semi-infinite Toda lattice and semi-infinite Pfaff 
    lattice are of use. Here we relate the semi-infinite Pfaff lattice of $B$-type to the semi-infinite Toda lattice.

    \paragraph{Semi-infinite Toda lattice.}  First, let us recall that the Hirota equation for the Toda lattice tau function 
    \cite{JM}, \cite{UT} and for the two-component KP tau function \cite{JM} coincide up to a sign factor, 
    see the relation (9.7) in \cite{JM} and the Theorem 1.12 in \cite{UT}. 
    Here  we shall consider the semi-infinite Toda lattice which may be presented as 
     \be\label{TL=cases2KP'}
 \tau^{\rm TL}_n(t,{\bar t})= (-)^{\frac{n(n+1)}{2}}\tau^{\rm 2KP}_n(t,{\bar t}) 
 \ee
    \be\label{semiinfiteTLtau}
    \tau^{\rm 2KP}_N(t^{(1)},t^{(2)}) = \l N,-N | e^{\sum_{a=1,2}\sum_{i\in\mathbb{Z}} \alpha^{(a)}_i t_i^{(a)}}
    e^{\sum_{i,j} M_{ij} \psi^{(1)}_i\psi^{\dag(2)}_j} |0\r 
    \ee
    (see also Appendix \ref{TL-2-KP}).
    
    The tau function (\ref{semiinfiteTLtau}) is the tau function $\tau_{N,0,0}^{(0)}$ of (\ref{two-component-TL-tau}) where 
    $g=e^{\sum_{i,j} M_{ij} \psi^{(1)}_i\psi^{\dag(2)}_j}$. This choice provides semi-infinity of the TL
    equation which is
    \be\label{TL-Hirota-equation}
    \frac{\partial^2 \tau^{\rm TL}_N}{\partial t^{(1)}_1\partial { t}^{(2)}_-1} \tau^{\rm TL}_N -
    \frac{\partial \tau^{\rm TL}_N}{\partial t^{(1)}_1}\frac{\partial\tau^{\rm TL}_N}{ \partial {t}^{(2)}_1} =
    - \tau^{\rm TL}_{N+1} \tau^{\rm TL}_{N-1}
    \ee   
    where we put $ \tau^{\rm TL}_N(t^{(1)},t^{(2)})=\delta_{N,0}$ for $N\le 0$. Given
    $ \tau^{\rm TL}_1$ one can construct all $ \tau^{\rm TL}_N\,,N>1$ in a recurrent way via (\ref{TL-Hirota-equation}).

    Introduce
    \[
     e^{\sum_{k>0} z^k t_k}=:\sum_{k\ge 0} z^k s_k(t) \,,\quad
     s_{\{h\}}(t):=\det \left( s_{h_i-j} \right)_{i,j=1,\dots,N}
    \]
    where $h=(h_1,\dots,h_N)$, $h_1>\cdots > h_N\ge 0$, and $s_{\{h\}}$ is the Schur function related
    to the partition $\lambda=(\lambda_1,\cdots,\lambda_N)$, $\lambda_i=h_i+i-N$, see \cite{Mac}.

    Using the relations 
    \[
    e^{\sum_{i\in\mathbb{Z}} \alpha^{(a)}_i t_i^{(a)}} \psi_j^{(a)}
    e^{-\sum_{i\in\mathbb{Z}} \alpha^{(a)}_i t_i^{(a)}}=\sum_{k\ge 0} \psi_{j-k}^{(a)}s_k(t)
    \]
    \[
     e^{\sum_{i\in\mathbb{Z}} \alpha^{(a)}_i t_i^{(a)}} \psi_j^{\dag(a)}
    e^{-\sum_{i\in\mathbb{Z}} \alpha^{(a)}_i t_i^{(a)}}=\sum_{k\ge 0} \psi_{j+k}^{\dag(a)} s_k(-t)
    \]
    and Wick's rule (see Appendix \ref{Pfaffians}), the tau function (\ref{semiinfiteTLtau}) may be presented 
    in the determinantal form:      
    \be\label{TL-tau-det}
    \tau^{\rm TL}_N(t^{(1)},t^{(2)})=\det \left( \textbf{m}_{ij}(t^{(1)},t^{(2)} \right)_{i,j=0,\dots,N-1}
    \ee
    where 
    \be\label{m-quasi-sym}
    \textbf{m}_{ij}(t^{(1)},t^{(2)})= \sum_{k,l \ge 0} M_{i+k,j+l} s_{k}(t^{(1)})s_{l}(-t^{(2)})
    \ee    
    As we see $\tau^{\rm TL}_1=\textbf{m}_{00}$.
    
    The tau function (\ref{semiinfiteTLtau}) may  also  be written in Takasaki form \cite{Takasaki-Schur},\cite{TI},\cite{TII}
    as
    \be\label{TL-Schur}
    \tau^{\rm TL}_N(t^{(1)},t^{(2)})=
    \sum_{h_1>\cdots > h_N \ge 0\atop h'_1>\cdots > h'_N \ge 0} M_{h,h'} s_{\{h\}}(t^{(1)})s_{\{h'\}}(-t^{(2)})
    \ee
    where
     \be\label{M-hh'}
    M_{\{h,h'\}} : =  \det\left[M_{h_i,h'_j} \right]_{i,j=1}^N
     \ee
   Series (\ref{TL-Schur}), where $M$ is specified, appear in various problems of random matrices, random partitions
   and in counting problems.
   
   Another way to present tau functions of the semi-infinite TL tau functions (which is of use in many models of 
   random matrices, see \cite{HO-2007} and Examples below) is
    \be\label{semiinfiteTLtau-int}
    \tau^{\it{TL}}_{n,m}(t^{(1)},t^{(2)},{\bar t}^{(1)},{\bar t}^{(2)}) = 
    \l n,-n | e^{\sum_{a=1,2}\sum_{i\in\mathbb{Z}} \alpha^{(a)}_i t_i^{(a)}}
    e^{\int  \psi^{(1)}(z_1)\psi^{\dag(2)}(z_2) d\mu(z_1,z_2)} 
    e^{\sum_{a=1,2}\sum_{i\in\mathbb{Z}} \alpha^{(a)}_{-i} {\bar t}_i^{(a)}}|m,-m\r 
    \ee 
    where $d\mu(z_1,z_2)$ is a bi-measure which should be specified according to a problem we are interested in.
    There are additional parameters here, ${\bar t}^{(1)},{\bar t}^{(2)}$ and $m$, which are hidden parameters of
    Toda lattice solutions. This is a particular case of the tau function (\ref{two-component-TL-tau}).
   
   From (\ref{semiinfiteTLtau-int}) we obtain  
   $\tau^{\it{TL}}_{n,m}(t^{(1)},t^{(2)},{\bar t}^{(1)},{\bar t}^{(2)})=\delta_{n,m}$ for $n\le m$,
   and
   \be\label{m+1,m}
   \tau^{\it{TL}}_{m+1,m}(t^{(1)},t^{(2)},{\bar t}^{(1)},{\bar t}^{(2)})= c(t)\int z_1^{m} z_2^{m}
   e^{\sum_{i\ge 1}\left(z_1^i t_i^{(1)} - z_2^i t_i^{(2)} - z_1^{-i} {\bar t}_i^{(1)}+z_2^{-i} {\bar t}_i^{(2)}  \right) }
   d\mu(z_1,z_2)
   \ee
   \[
    c(t)=
    e^{\sum_{i\ge 1}\left( it_i^{(1)}{\bar t}_i^{(1)} + i t_i^{(2)}{\bar t}_i^{(1)}  \right) }
   \]

    {\bf Example 1.1}.  The choice $d\mu(z_1,z_2)=e^{-|z|^2}\delta^{(2)}(z_2-{\bar z}_1)d^2z_1 d^2z_2  $, 
    $z_{1,2} \in\mathbb{C}$ yields (see \cite{HO-2007}) both the fermionic representation for the partition function
    of the ensemble of normal matrices (about this ensemble see \cite{CZ} and \cite{MWZ}), and the partition function
    of the complex Ginibre ensemble (about Ginibre ensemble of complex matrices see Ch. 15.1 in \cite{Mehta}). In this 
    example ${\bar z}$ is the complex conjugate of $z$. Here $\delta^{(2)}$ is the two-dimensional delta function.
    
     {\bf Example 1.2}. The choice $d\mu(z_1,z_2)= e^{z_1 z_2}dz_1 dz_2  $, 
    $z_{1,2} \in\mathbb{R}$ yields (see \cite{HO-2007}) the fermionic representation for the partition function
    of the two-matrix model introduced in \cite{IZ} (the relation of the two-matrix model and TL see in \cite{GMMMO} and 
    different fermionic representation see in \cite{KMMOZ}).  In case we take $z_{1,2} \in S^1$ we obtain the model
    of two unitary matrices \cite{ZinnJustin}. 
    
     {\bf Example 1.3}. The choice  $d\mu(z_1,z_2)= \delta(z_1-z_2) dz_1 dz_2  $, 
    $z_{1,2} \in\mathbb{R}$ yields (see \cite{Mironych-2-komp}) the fermionic representation for the partition function
    of the one-matrix model (the relation of the one-matrix model and one-dimensional TL see in \cite{GMMMO} and 
    different fermionic representation see in \cite{KMMOZ}).
    
     {\bf Example 1.4}.  Take $z_a=e^{i\phi_a}\,,a=1,2$. The choice  $d\mu(e^{i\phi_1},e^{i\phi_2})=\delta(\phi_1-\phi_2)
    d\phi_1 d\phi_2$, $0\le   \phi_1,\phi_2,\phi \le 2\pi$, 
    yields the fermionic representation for the $\beta=2$ circlar ensemble (about this ensemble
    see Ch. 10.3 in \cite{Mehta} and \cite{MMS}).

    In case the bi-measure $d\mu$ and the matrix $M$ are related by the moment's transform
    \be
    M_{ij}=M_{ij}({\bar t}^{(1)},{\bar t}^{(2)},m)=   \int z^{i+m}_1 z^{j+m}_2 
    e^{\sum_{k\ge 0} \left({\bar t}_k^{(1)} z_1^{-k} -{\bar t}_k^{(2)} z_2^{-k} \right)} d\mu(z_1,z_2)
    \ee
   then the tau functions (\ref{semiinfiteTLtau}) and (\ref{semiinfiteTLtau}) may be equated  as follows
   \be
   \tau^{\rm 2KP}_N(t^{(1)},t^{(2)};{\bar t}^{(1)},{\bar t}^{(2)},m)= 
   e^{\sum_{a=1,2}\sum_{k\ge 1} kt^{(a)}_k {\bar t}^{(a)}_k }\tau^{TL}_{N+m,m}(t^{(1)},t^{(2)},{\bar t}^{(1)},{\bar t}^{(2)})
   \ee
   Now the tau function (\ref{semiinfiteTLtau}) written here depends on the additional parameters 
   ${\bar t}^{(1)},{\bar t}^{(2)},m$. Expression for $\tau^{\rm TL}_1$ given by (\ref{m+1,m}) allows to obtain
   all  $\tau^{\rm TL}_N\,,N>1$, in the reccurent way from TL Hirota equation (\ref{TL-Hirota-equation}).

    \paragraph{Semi-infinite TL and semi-infinite B-type Pfaff Lattice (BPL).}
    \bp\label{BPF^2=TL}
    \be
    \left(\tau^B_N(t)\right)^2 = \tau^{\rm TL}_N(t,-t)
    \ee
    where
    \be\label{realization-via-sum}
    \tau^B_N(t)=\l N| e^{\sum_{i\in\mathbb{Z}} \beta_i t_i}
  e^{\frac12\sum_{i,j} A_{ij} \psi_i\psi_j +  \sqrt{2}\sum_{i} a_i\psi_i \varphi } 
  e^{\sum_{i\in\mathbb{Z}} \beta_{-i} {\bar t}_i}|0\r
    \ee
    \be
    \tau^{\rm TL}_N(t,-t)=(-)^{\frac{N(N+1)}{2}}
    \l N,-N| e^{\sum_{i\in\mathbb{Z}} \left( \alpha^{(1)}_i-\alpha^{(2)}_i\right) t_i }
  e^{\sum_{i,j} \left(A_{ij}-a_ia_j \right) \psi^{(1)}_i\psi^{\dag(2)}} |0\r
    \ee
  and  $A_{ij}=-A_{ji}$.    
    \ep

    It is well-known that the determinant of a skew symmetric $N\times N$ matrix vanishes if $N$ is odd.
    The  square root of a skew symmetric matrix of even size is the Pfaffian of this matrix. 
    Let us call the sum of a skew symmetric matrix and a symmetric matrix of rank 1 the {\em quasi-skew symmetric} matrix.
    The square root of a quasi-skew symmetric matrix may be
    identified with a Pfaffian of some different matrix:
 \bl\label{Pf-quasi-skew-det} {\em The square root of the determinant of the quasi-skew symmetric $N \times N$ matrix with 
 entries $A_{ij}-a_ia_j$, $i,j=1,\dots, N$, where $A_{ij}=-A_{ji}$,
 is the pfaffian of the $2\left[\frac {N+1}{2} \right] \times 2\left[\frac {N+1}{2} \right]  $ 
 matrix $B$ which is defined as follows. For $N$ even, $B=A$. For $N$
 odd, $N=2n-1$,
 \be \label{A-alpha-odd-n} 
 {\textsc{B}}_{ij}=-{\textsc{B}}_{ji}:=
\cases {
A_{i,j} \quad \mbox{ if }\quad 1\le i<j \le 2n-1 \cr
a_{i} \qquad \, \mbox{ if }\quad 1\le i < j=2n } .
  \ee

Thus, for both for odd and for even $N$, we have
\be\label{pfaff=sqrt-det-tilde-A}
\left( \Pf B \right)^2 =  \det\left(A_{i,j} - a_{i}a_{j}  \right)_{i,j=1,\dots,N}
\ee
}
 \el
    
We have the following 
 \bc
 \be
 \left( \sum_{h_1>\cdots >h_N} \sqrt {\det(A_{h_i,h_j}-a_{h_i}a_{h_j})} s_{\{h\}}(t) \right)^2=
 \sum_{h_1>\cdots >h_N \atop h'_1>\cdots >h'_N} \det(A_{h_i,h'_j}-a_{h_i}a_{h'_j}) s_{\{h\}}(t)s_{\{h'\}}(t)
 \ee
 \ec

  \bp\label{BPF^2=TL-two-sided}
    \be
    \left(\tau^B_{n,m}(t,{\bar t})\right)^2 = 
    \tau^{ TL}_{n,m}(t,-t;{\bar t},-{\bar t})
    \ee
    where
    \be\label{realization-via-integral}
    \tau^B_{n,m} (t,{\bar t})  =\l n| e^{\sum_{i\in\mathbb{Z}} \beta_i t_i}
  e^{\frac12\int \psi(z_1)\psi(z_2) d\nu(z_1,z_2) + \sqrt{2}\int \psi(z) d\nu(z) \varphi } 
  e^{\sum_{i\in\mathbb{Z}} \beta_{-i} {\bar t}_i}|m\r
    \ee
    \be
    (-)^{\frac{n(n+1)}{2}+\frac{m(m+1)}{2}}\tau^{\rm TL}_{n,m}(t,-t;{\bar t},-{\bar t})=
    \ee
    \be
    \l n,-n| e^{\sum_{i\in\mathbb{Z}} \left( \alpha^{(1)}_i-\alpha^{(2)}_i\right) t_i }
  e^{\int \psi^{(1)}(z_1)\psi^{\dag(2)}(z_2) \left(d\omega(z_1,z_2)-d\nu(z_1)d\nu(z_2) \right) }
  e^{\sum_{i\in\mathbb{Z}}\left( \alpha^{(1)}_{-i}-\alpha^{(2)}_{-i} \right){\bar t}_i } |m,-m\r
    \ee
  and where  $d\omega(z_1,z_2)=-d\omega(z_1,z_2)$ and $d\nu(z)$ are measures.     
    \ep

    Examples of the partition functions of various ensembles and also of some useful multiple integrals
    obtained as the semi-infinite BPL tau functions:
    
    {\bf Example 2.1}. The choice $d\omega(z_1,z_2)=\frac12\sgn(z_1-z_2)dz_1 dz_2$ and $d\nu(z)=dz$, $z_{1,2},z \in\mathbb{R}$, 
    yields the fermionic representation for the $\beta=1$ ensemble (orthogonal ensemble, see Ch. 7 in \cite{Mehta}) \cite{L1}.
    
    {\bf Example 2.2}. The choice $d\omega(z_1,z_2)=\frac14\left(\partial_{z_1}- \partial_{z_2}\right)\delta(z_1-z_2)dz_1 dz_2$ 
    and $d\nu(z)=0$ , $z_{1,2} \in\mathbb{R}$,
    yields the fermionic representation for the $\beta=4$ ensemble (symplectic ensemble, see Ch. 8 in \cite{Mehta}) \cite{L1}.
    
    {\bf Example 2.3}. Take $z_a=e^{i\phi_a}$. The choice  $d\omega(e^{i\phi_1},e^{i\phi_2})=
    \frac12\sgn(\phi_1- \phi_2)d\phi_1 d\phi_2$ and 
    $d\nu(e^{i\phi})=d\phi$, $0\le   \phi_1,\phi_2,\phi \le 2\pi$, 
    yields (see \cite{OST-II}) the fermionic representation for the $\beta=1$ circular ensemble (about this ensemble
    see Ch. 10.1 in \cite{Mehta}).
    
    {\bf Example 2.4}. The choice  $d\omega(e^{i\phi_1},e^{i\phi_2})=
    \frac14\left(\partial_{\phi_1}- \partial_{\phi_2}\right)\delta(\phi_1- \phi_2)d\phi_1 d\phi_2$ and 
    $d\nu(e^{i\phi})=d\phi$, $0\le   \phi_1,\phi_2,\phi \le 2\pi$, 
    yields (see \cite{OST-II}) the fermionic representation for the $\beta=4$ circular ensemble (about this ensemble
    see Ch. 10.2 in \cite{Mehta}).    
    
    {\bf Example 2.5}. The choice $d\omega(z_1,z_2)=\delta(z_1+z_2)dz_1 dz_2$ and $d\nu(z)=\delta(z)dz$, 
    $z_{1,2},z \in\mathbb{R}$,    yields (see \cite{OST-II}) the fermionic representation for the ensemble  of 
    antisymmetric matrices (see Ch. 13 in \cite{Mehta} about this ensemble) .
    
    {\bf Example 2.6}. For the Pandey-Mehta interpolating ensembles, $0< \alpha^2<\infty $ being the interpolation 
    parameter, see \cite{Mehta} Ch. 14
    
    {\bf (a)} for the ensemble interpolating between GUE  and GOE, see \cite{Mehta} Ch. 14.1,  we take 
   \[
   d\omega(z_1,z_2)= e^{-\frac12 (1+\alpha^2)(z_1^2+z_2^2)} 
   \,{\rm erf}\left(\sqrt{\frac{1-\alpha^4}{4\alpha^2}}(z_1-z_2)\right)dz_1 dz_2,\quad 
   d\nu(z)= e^{-\frac12 (1+\alpha^2)z^2} dz,\quad
   z_{1,2},z \in\mathbb{R}\] 
   Here ${\rm erf}(x)=\int_0^x e^{-x^2}dx$. 
   
   {\bf (b)} for the ensemble interpolating between GUE  and GSE, see \cite{Mehta} Ch. 14.2,  we take 
   \[
   d\omega(z_1,z_2)= (z_1-z_2)e^{-\frac12 (1+\alpha^2)(z_1^2+z_2^2)} 
   \,e^{-\frac{(1-\alpha^4)(z_1-z_2)^2}{4\alpha^2}}dz_1 dz_2,\quad 
   z_{1,2} \in\mathbb{R}
   \] 
     
    The same measures $d\omega$, $d\nu$ are available to get the asymmetric two-matrix model where the first matrix is 
    Hermitian while the second is respectively (a) either symmetric ($0<\alpha^2 < 1$), or  antisymmetric($1<\alpha^2 < \infty$)  
    (b) either self-dual  ($0<\alpha^2 < 1$),   or anti-self-dual ($1 <\alpha^2 < \infty$), see \cite{OST-II}
  
    {\bf Example 2.7}. The choice $d\omega(z_1,z_2)=\frac12({\bar z}-z)e^{-|z|^2}\delta^{(2)}(z_2-{\bar z}_1)d^2z_1 d^2z_2  $, 
    $z_{1,2} \in\mathbb{C}$ and $d\nu(z)=0$ yields (see \cite{O-2012})
    the fermionic representation for the ensemble of random (non-Hermitian) quaternionic matrices (the so-called quaternion-real 
    Ginibre ensemble (see Ch. 15.2 in \cite{Mehta}) . Here $\delta^{(2)}$ is the two-dimensional delta function. In this example
    ${\bar z}$ is the complex conjugate of $z$.
    
    {\bf Example 2.8}. For the Ginibre ensemble of real matrices (see Ch. 15.3 in \cite{Mehta}) the choice of the measure is more 
    complicated, and may be found in \cite{O-2012}.
  
    {\bf Example 2.9}.  The choice $d\omega(z_1,z_2)=\frac12\frac{z_1-z_2}{z_1+z_2}dz_1 dz_2$, $z_{1,2} \in\mathbb{R}$,
    yields (see \cite{OST-II}) the fermionic representation for the so-called Bures ensembles which appears in quantum 
    chaos problems where random density matrix appears \cite{OsipovSommers}. 
    
    {\bf Example 2.10}.   The choice $d\omega(z_1,z_2)=\frac12 \tanh \pi (z_1-z_2)  dz_1 dz_2$, $z_{1,2} \in\mathbb{R}$,
    yields (see \cite{OST-II}) the fermionic representation for the Plancheral measure for the group $SL(N,\mathbb{R})$ 
    (and for the symmetric space $SL(N,\mathbb{R})/SO(N)$),  see Section 17.2.8 in \cite{VKIII} where put 
    $\lambda_i=\frac 12 z_i$. 
       
    {\bf Examples 2.11}. The choice $d\omega(z_1,z_2)=\frac{1}{2\pi i}\left(z_1-z_1^{-1} \right)\delta(z_2-z_1^{-1})dz_1 dz_2$ 
     and $d\nu(z)=0$, $z_{1,2}, z \in S^{1}$ yields the ensemble of symplectic matrices, details see in the forthcoming
     paper \cite{LO-preprints}
       
    {\bf Examples 2.12}. The choice $d\omega(z_1,z_2)=\frac{1}{2\pi i}\left(z_2^{-1}-z_2 \right)^{-1}\delta(z_2-z_1^{-1})dz_1 dz_2$
    and $d\nu(z)=\frac12\delta(z-1)dz$, $z_{1,2},z \in S^{1}$ yields the ensemble of orthogonal matrices, details see in the 
    forthcoming paper \cite{LO-preprints}.

  The relation between realizations (\ref{realization-via-sum}) and (\ref{realization-via-integral})
  is given by
   \be
  A_{ij} = A_{ij} ({\bar t},m) = \int z_1^{i+m} z_2^{j+m} 
   e^{\sum_{k\ge 0} {\bar t}_k \left(z_1^{-k}+ z_2^{-k} \right)} d\nu(z_1,z_2)\,,\quad
  a_i = a_i({\bar t},m)= \int z^{i+m}  
   e^{\sum_{k\ge 0} {\bar t}_k z^{-k}} d\nu(z)
   \ee

   In $B$-case relations (\ref{TL-tau-det}),(\ref{m-quasi-sym}) read as
   \be\label{TL-tau-det-B}
    \tau^{\rm TL}_N(t,-t;{\bar t},-{\bar t},m)=\det \left( \textbf{m}_{ij}(t;{\bar t},m) \right)_{i,j=0,\dots ,N-1}
   \ee
   \be\label{m-quasi-sym-B}
    \textbf{m}_{ij}(t;{\bar t},m)= 
    \int z_1^{i+m}z_2^{j+m} e^{\sum_{k\ge 1}(z_1^k+z_2^k)t_k -(z_1^{-k}+z_2^{-k}){\bar t}_k}
    \left( d\omega(z_1,z_2)- d\nu(z_1)d\nu(z_1) \right)
   \ee
   \be
   =   \sum_{k,l \ge 0} A_{i+k,j+l}({\bar t},m) s_{k}(t)s_{l}(t) -
    \sum_{k\ge 0} a_{i+k}({\bar t},m)s_{k}(t)\sum_{l\ge 0} a_{j+l}({\bar t},m) s_{l}(t)
   \ee
   As we see the matrix $\textbf{m}$ is quasi-skew symmetric one.
   
   Then for both odd and even $N$ we have from Proposition \ref{BPF^2=TL-two-sided}
   \be\label{BPL}
    \left(\tau^{B}_N(t;{\bar t},m)\right)^2 = 
    \det \left( \textbf{m}_{ij}(t;{\bar t},m) \right)_{i,j=0,\dots,N-1}
   \ee
 According to Lemma \ref{Pf-quasi-skew-det}, $\tau^{B}_N$ is a certain Pfaffian (this result may be obtained also from
 the Wick theorem applied to the fermionic expectation value (\ref{realization-via-integral}).
 
   We obtain a generalization of the result presented in \cite{AvM-Pfaff}, \cite{AMS} where the relation
(\ref{BPL}) was achieved  for the case of even $N$ and for skew symmetric matrices $\textbf{m}$. This case
may referred as the $D$-type Pfaff lattice (DPL), see \cite{T-09}.

\section{Outlook}

In our next paper \cite{LO-preprints} the analogue of the Proposition \ref{Pfaff-det-tau} for the multicomponent BKP tau functions \cite{KvdLbispec} will be written down.
Also we shall consider the relations between various matrix integrals and between series over partitions which result from Propositions \ref{BPF^2=TL-two-sided}
and \ref{Pfaff-det-tau}.

\section*{Acknowledgements}

One of the authors (A.O.) thanks K.Takasaki and T.Shiota for helpful discussions regarding Toda lattice and Pfaff lattice.
Не thanks RFBR grant 14-01-00860.

\appendix

\section{Appendices}

\subsection{Hirota equation for the TL and for the 2-component KP tau functions. \label{TL-2-KP}} 
 The TL tau function was introduced in \cite{JM} 
 and may be defined by 
  \be
  \tau^{\rm TL}_n(t,{\bar t}) = \l n|e^{\sum_{i>0} t_i\alpha_i} g^{\rm TL} e^{-\sum_{i>0} {\bar t}_i\alpha_{-i}} |n\r
  \ee

This tau function solves Hirota equation, \cite{JM},\cite{UT}
 \bea\label{Hirota-TL-}
 \oint\frac{dz}{2\pi i}z^{n'-n}e^{V(t'-t,z)}\tau^{\rm TL}_{n'}\left(t'-[z^{-1}],{\bar t}'  \right)
 \tau^{\rm TL}_{n}\left(t+[z^{-1}],{\bar t}  \right)=   \nonumber \\  \oint\frac{dz}{2\pi i} 
 z^{n'-n}e^{V\left({\bar t}'-{\bar t},z^{-1}\right)}\tau^{\rm TL}_{n'+1}\left(t',{\bar t}'-[z]  \right)
 \tau^{\rm TL}_{n-1}\left(t,{\bar t} +[z] \right) 
 \eea
(see \cite{JM}, \cite{UT}) which includes
\be\label{the-first-TL-Hirota}
\frac{\partial^2 \tau^{\rm TL}_n}{\partial t_1\partial {\bar t}_1} \tau^{\rm TL}_n -
\frac{\partial \tau^{\rm TL}_n}{\partial t_1}\frac{\partial\tau^{\rm TL}_n}{ \partial {\bar t}_1} =
-\tau^{\rm TL}_{n+1} \tau^{\rm TL}_{n-1}
\ee

  The two-component KP  tau function
 \be
 \tau^{\rm 2KP}_n(t,{\bar t}) =
 \l n,-n| e^{\sum_{i>0} \left(t_i'\alpha_i^{(1)} +{\bar t}_i'\alpha_i^{(2)}\right)} g^{\rm 2KP} |0\r
 \ee
 solves Hirota equation 
  \bea\label{Hirota-2KP-derivation}
 \oint\frac{dz}{2\pi i}(-)^{-n'-n}z^{n'-n}e^{V(t'-t,z)}\tau^{\rm 2KP}_{n'}\left(t'-[z^{-1}],{\bar t}'  \right)
 \tau^{\rm 2KP}_{n}\left(t+[z^{-1}],{\bar t}  \right) =   \nonumber \\  \oint\frac{dz}{2\pi i} 
 z^{n-n'-2}e^{V\left({\bar t}'-{\bar t},z\right)}\tau^{\rm 2KP}_{n'+1}\left(t',{\bar t}'-[z^{-1}]  \right)
 \tau^{\rm 2KP}_{n-1}\left(t,{\bar t} +[z^{-1}] \right)
 \eea
 which up to the sign factor $(-)^{n+n'}$ in the first integral is (\ref{Hirota-TL-}) if we change $z\to z^{-1}$ 
 in the second integral in (\ref{Hirota-2KP-derivation}). This brings us to the relation (\ref{TL=cases2KP'}).

 \subsection{Hirota equation for the two-sided two-component KP. \label{two-sided-Hirota}}

 The two-sided two-component KP tau function (\ref{two-component-TL-tau}) 
  \[
  \tau_{N}\left(L^{(1)}, {L^{(2)}};t^{(1)},t^{(2)};{\bar t}^{(1)},{\bar t}^{(1)}\right)=
  \]
  \be\label{tau-two-sided-two-component-KP}
  \l N+ L^{(1)},-N+ L^{(2)}\vert 
  e^{ \sum_{k=1}^\infty \left( t^{(1)}_k\alpha^{(1)}_k + t^{(2)}_k\alpha^{(1)}_k \right) }  \, g\,
  e^{ \sum_{k=1}^\infty \left( {\bar t}^{(1)}_{-k}\alpha^{(1)}_{-k} + {\bar t}^{(2)}_k\alpha^{(2)}_{-k} \right) }
  \vert L^{(1)}, L^{(2)}\r
  \ee
solves the following  Hirota equations 
\be\label{Hirota-two-sided-2component-KP}
  \oint\frac{dz}{2\pi i} {\rm Bil}\left(N',{\bf L}';{\bf t}';{\bar {\bf t}}'; N,{\bf L};{\bf t};{\bar {\bf t}};z\right)=0\,,
 \ee
 \[
 {\rm Bil}\left(N',{\bf L}';{\bf t}';{\bar {\bf t}}'; N,{\bf L};{\bf t};{\bar {\bf t}};z\right)\,:=
 \]
  \bea
  (-z)^{N'-N+{L^{(1)}{}'}-L^{(1)}}e^{V\left({t^{(1)}{}'}-{t^{(1)}},z\right)}
  \tau^{(0)}_{N'}\left({L^{(1)} {}'}, {L^{(2)}{ }'};{t_-^{(1)}{}'}(z),{{t^{(2)}{}'}};{\bar {\bf t}{}'}\right)
  \tau^{(0)}_{N}\left(L^{(1)}, {L^{(2)}};t_+^{(1)}(z),t^{(2)};{\bar {\bf t}}\right) \nonumber\\
+ z^{N-N'+{{L^{(2)}}'}-L^{(2)}-2}e^{V\left({t^{(2)}{}'}-{t^{(2)}},z\right)}
  \tau^{(0)}_{N'+1}\left({L^{(1)}{}'},{L^{(2)}{}'};{t^{(1)}{}'},{t_-^{(2)}{}'}(z);{{\bar {\bf t}}{}'}\right)
  \tau^{(0)}_{N-1}\left(L^{(1)}, {L^{(2)}};t^{(1)},{t_+^{(2)}}(z);{\bar{\bf t}}\right) \nonumber\\
- (-z)^{{L^{(1)}{}'}-L^{(1)}}
e^{V\left({\bar t}^{(1)}- {{\bar t}^{(1)}{}'},z^{-1}\right)} 
  \tau^{(0)}_{N'}\left( {L_+^{(1)}{}'},{L^{(2)}{}'};{\bf t}',{{\bar t}_+^{(1)}{}'}(z),{{\bar t}^{(2)}{}'}\right)
  \tau^{(0)}_{N}\left(L_-^{(1)}, L^{(2)};{\bf t};{\bar t}^{(1)}_-(z),{\bar t}^{(2)}\right) \nonumber \\
  - z^{{L^{(2)}{}'} - L^{(2)}}e^{V\left({\bar t}^{(2)}-{{\bar t}^{(2)}{}'},z^{-1}\right)}
\tau^{(0)}_{N'+1}\left({L^{(1)}{}'},  {L_+^{(2)}{}'} ;{\bf t}';{{\bar t}^{(1)}{}'},{{\bar t}_+^{(2)}{}'}(z)\right)
  \tau^{(0)}_{N-1}\left(L^{(1)}, L_-^{(2)};{\bf t};{\bar t}^{(1)},{\bar t}^{(2)}_-(z)\right)  \nonumber \\
  \label{hirota-two-sided-2component-KP}
\eea
where ${\bf L}'=\left({L^{(1)}{}'},  {L^{(2)}{}'}\right)$ and ${\bf L}=\left (L^{(1)},  L^{(2)}\right)$ are pairs 
of  
integers, and
${\bf t}'=\left( {{\bar t}^{(1)}{}'},{{\bar t}^{(2)}{}'} \right)$, 
${\bf t}'=\left( {\bar t}^{(1)},{\bar t}^{(2)} \right)$,
${{\bar t}^{(a)}{}'}=\left( t_1^{(2)},  t_2^{(a)},\dots \right)$,
${{\bar t}^{(a)}{}'}=\left({{\bar t}_1^{(2)}{}'}, {{\bar t}_2^{(a)}{}'},\dots \right)$ are
semi-infinite sets of higher times. Then
\[
{t_\pm^{(a)}{}'}(z):={t^{(a)}{}'}\pm [z^{-1}]\,,\quad
{{\bar t}_\pm^{(a)}{}'}(z):= {{\bar t}^{(a)}{}'}\pm [z]\,,\quad
{t_\pm^{(a)}}(z):={t^{(a)}}\pm [z^{-1}]\,,\quad
{{\bar t}_\pm^{(a)}}(z):= {{\bar t}^{(a)}}\pm [z]\,,\quad
\]
and
\[
  L_\pm^{(a)}{}' = L^{(a)}{}'\pm 1\,,\quad  L^{(a)}_\pm = L^{(a)}\pm 1
\]
where $a=1,2$.

 \br
 
 By replacing $z\to z^{-1}$ in the last two members of (\ref{hirota-two-sided-2component-KP}) 
we obtain Hirota 
 equation for the 4-component KP (see \cite{JM} on multicomponent KP), where $t^{(3)}$ and $t^{(4)}$ may be 
 identified respectively with ${{\bar t}^{(1)}}$ and ${{\bar t}^{(1)}}$.

 The two-sided two-component KP is a particular case ($p=4$) of the $p$-component KP tau function, introduced 
 in \cite{JM},
 \be
 \tau({\bf N}; {\bf s}):=
 \l N^{(1)},\dots, N^{(p)} \vert e^{\sum_{a=1}^p\sum_{i>0} \alpha^{a} t^{(a)}} g^{(1,\dots,p)} \vert 0,0\r\,
 \ee
where $g^{(1,\dots,p)}$ solves
\be\label{fermionic-Hirota-p-KP}
\left[ g^{(1,\dots,p)}\otimes g^{(1,\dots,p)}, \sum_{a=1}^p\sum_{i\in\mathbb{Z}} \psi^{(a)}_i\otimes \psi^{\dag(a)}_i 
\right]=0
\ee
 From (\ref{fermionic-Hirota-p-KP}) the multicomponent KP Hirota equations are obtained \cite{JM}:
\be\label{Hirota-p-KP}
  \sum_{a=1}^p\oint\frac{dz}{2\pi i}(-)^{\kappa_a}z^{N^{(a)}{'}-N^{(a)}-2}e^{V(t^{(a)}{'}-t^{(a)},z)}
  \tau\left({\bf N}_-^{[a]}{'} ;{\bf t}_-^{[a]}{'}(z)\right)
  \tau\left({\bf N}_+^{[a]}  ; {\bf t}_+^{[a]}(z)\right) =0
\ee
where 
\[
{\bf N}_\pm^{[a]} :=\left(N^{(1)},\dots,N^{(a-1)},N^{(a)}\pm 1,N^{(a+1)},\dots,N^{(p)} \right)
\]
\[
{\bf t}_\pm ^{[a]}(z):= \left(t^{(1)},\dots,t^{(a-1)}, t^{(a)}\pm [z^{-1}],t^{(a+1)},\dots, t^{(p)} \right)
\]
\be
\kappa_a = N_p +\cdots + N_{a+1} + N'_p +\cdots + N'_{a+1}
\ee
 
In (\ref{Hirota-p-KP}), ${\bf N}=\left(N^{(1)},\dots,N^{(p)} \right)$ and 
${\bf N}{'}=\left(N^{(1)}{'},\dots,N^{(p)}{'} \right)$ are two independent sets of vacuum charges, while
$t^{(a)}=\left( t^{(a)}_1,t^{(a)}_2, t^{(a)}_3, \right)$ and
$t^{(a)}{'}=\left( t^{(a)}_1{'},t^{(a)}_2{'}, t^{(a)}_3{'}, \right)$, $a=1,\dots,p$, are two independent sets 
of the multicomponent KP higher times.

 \er

 \subsection{Hirota equations for the two-sided BKP.\label{two-sided BKP-Hirota}}
 Hirota equations for the large BKP hierarchy were written in \cite{KvdLbispec}. For the two-sided BKP hierarchy
 (2-BKP hierarchy)  (\ref{2BKP-tau}), Hirota equations  are as follows \cite{OST-I}
 \bea\label{Hirota-2lBKPtau}
  \oint\frac{dz}{2\pi i}z^{N'+L'-N-L-2}e^{V(s'-s,z)}
  \tau_{N'-1}(L',s'-[z^{-1}],{\bar s}')
  \tau_{N+1}(L,s+[z^{-1}],{\bar s}) \nonumber\\
+ \oint\frac{dz}{2\pi i}z^{N+L-N'-L'-2}e^{V(s-s',z)}
  \tau_{N'+1}(L',s'+[z^{-1}],{\bar s}')
  \tau_{N-1}(L,s-[z^{-1}],{\bar s}) \nonumber\\
= \oint\frac{dz}{2\pi i}z^{L'-L}e^{V({\bar s}'-{\bar s},z^{-1})} 
  \tau_{N'-1}(L'+1,s',{\bar s}'-[z])
  \tau_{N+1}(L-1,s,{\bar s}-[z]) \nonumber \\
+ \oint\frac{dz}{2\pi i}z^{L-L'}e^{V({\bar s}'-{\bar s},z^{-1})}
  \tau_{N'+1}(L'-1,s',{\bar s}'+[z])
  \tau_{N-1}(L+1,s,{\bar s}+[z]) \nonumber\\
+ \frac{(-1)^{L'+L}}{2}(1-(-1)^{N'+N})
  \tau_{N'}(L',s',{\bar s}')\tau_N(L,s,{\bar s}) 
\eea
The difference BKP Hirota equation may be obtained from the previous one, see \cite{OST-I}
\bea
  - \frac{\beta}{\alpha-\beta}
    \tau_N(l,\bt+[\beta^{-1}])\tau_{N+1}(l,\bt+[\alpha^{-1}]) 
  - \frac{\alpha}{\beta-\alpha}
    \tau_N(l,\bt+[\alpha^{-1}])\tau_{N+1}(l,\bt+[\beta^{-1}]) 
  \nonumber\\
  + \frac{1}{\alpha\beta}
    \tau_{N+2}(l,\bt+[\alpha^{-1}]+[\beta^{-1}])\tau_{N-1}(l,\bt) 
  = \tau_{N+1}(l,\bt+[\alpha^{-1}]+[\beta^{-1}])\tau_N(l,\bt). 
\eea

By replacing $z\to z^{-1}$ in the last two members of (\ref{Hirota-2lBKPtau}) it may be written as follows 
\bea\label{Hirota-2lBKPtau-}
  \oint\frac{dz}{2\pi i}z^{N'+L'-N-L-2}e^{V(s'-s,z)}
  \tau_{N'-1}(L',s'-[z^{-1}],{\bar s}')
  \tau_{N+1}(L,s+[z^{-1}],{\bar s}) \nonumber\\
+ \oint\frac{dz}{2\pi i}z^{N+L-N'-L'-2}e^{V(s-s',z)}
  \tau_{N'+1}(L',s'+[z^{-1}],{\bar s}')
  \tau_{N-1}(L,s-[z^{-1}],{\bar s}) \nonumber\\
- \oint\frac{dz}{2\pi i}z^{L-L'-2}e^{V({\bar s}'-{\bar s},z)} 
  \tau_{N'-1}(L'+1,s',{\bar s}'-[z^{-1}])
  \tau_{N+1}(L-1,s,{\bar s}-[z^{-1}]) \nonumber \\
- \oint\frac{dz}{2\pi i}z^{L'-L-2}e^{V({\bar s}'-{\bar s},z)}
  \tau_{N'+1}(L'-1,s',{\bar s}'+[z^{-1}])
  \tau_{N-1}(L+1,s,{\bar s}+[z^{-1}]) \nonumber\\
= \frac{(-1)^{L'+L}}{2}(1-(-1)^{N'+N})
  \tau_{N'}(L',s',{\bar s}')\tau_N(L,s,{\bar s}) 
\eea
which is up to the sign factor the Hirota equation for the 2-component BKP \cite{KvdLbispec} tau function:
 \be
 \tau(N^{(1)},N^{(2)}; s^{(1)},s^{(2)}):= (-)^{\frac{N^{(2)}(N^{(2)}+1)}{2}}
 \l N^{(1)},N^{(2)} \vert e^{\sum_{a=1,2}\sum_{i>0} \beta^{a} s^{(a)}} h^{(1,2)} \vert 0,0\r
 \ee

 \subsection{Pfaffians, the Wick's rule \label{Pfaffians}}

\paragraph{Pfaffian.} If $A$ an anti-symmetric matrix of an odd order its determinant
vanishes. For even order, say $k$, the following multilinear form
in $A_{ij},i<j\le k$
 \be\label{Pf''}
\Pf [A] :=\sum_\sigma
{\sgn(\sigma)}\,A_{\sigma(1),\sigma(2)}A_{\sigma(3),\sigma(4)}\cdots
A_{\sigma(k-1),\sigma(k)}
 \ee
where sum runs over all permutation restricted by
 \be
\sigma:\,\sigma(2i-1)<\sigma(2i),\quad\sigma(1)<\sigma(3)<\cdots<\sigma(k-1),
 \ee
 coincides with the square root of $\det A$ and is called the
 {\em Pfaffian} of $A$, see, for instance \cite{Mehta}. 
 
 \paragraph{ Wick's relations.} Let each of $w_i$ be a linear
combination of Fermi operators:
  \[
{\hat w}_i=\sum_{m\in\mathbb{Z}}\,v_{im}\psi_m\,+\,
\sum_{m\in\mathbb{Z}}\,u_{im}\psi^\dag_m\,
 ,\quad i=1,\dots,n
 \]
  Then the Wick formula is
  \be\label{Wick} \l l|{\hat w}_1\cdots {\hat w}_n |l\r =
\cases{
\Pf\left[ A \right]_{i,j=1,\dots,n} \quad {\rm if\,\,n\,is\,even}  \cr
0 \qquad\qquad\qquad\quad\, \mbox{otherwise}
 }
  \ee  
    where $A$ is $n$ by $n$ antisymmetric matrix with entries
 $ A_{ij}\, = \,\l l|{\hat w}_i {\hat w}_j|l\r\, ,\quad i<j$.

\end{document}